\documentclass[
    twocolumn,
    superscriptaddress,
    amsmath,
    amssymb,
    aps,
    prb,
    floatfix
]{revtex4-2}

\usepackage{graphicx}
\usepackage{dcolumn}
\usepackage{bm}

\usepackage{romannum}
\usepackage{hyperref}
\usepackage{comment}
\usepackage{float}
\usepackage{physics}
\usepackage[caption=false]{subfig}
\usepackage{xspace}
\usepackage{amsbsy}
\usepackage[utf8]{inputenc}
\usepackage{siunitx}
\usepackage{multirow}

\hypersetup{
    colorlinks = true,
    allcolors = {blue},
}

\begin{document}

\title{Generation of gigahertz frequency surface acoustic waves in YIG/ZnO heterostructures}%

\author{Finlay Ryburn}
\email{finlay.ryburn@physics.ox.ac.uk}
\affiliation{Clarendon Laboratory, Department of Physics, University of Oxford, Parks Road, Oxford, OX1\,3PU, United Kingdom}
\author{Kevin Künstle}
\email{kuenstle@rptu.de}
\affiliation{Fachbereich Physik and Landesforschungszentrum OPTIMAS, Rheinland-Pfälzische Technische Universität Kaiserslautern-Landau, 67663 Kaiserslautern, Germany}
\author{Yangzhan Zhang}
\affiliation{Clarendon Laboratory, Department of Physics, University of Oxford, Parks Road, Oxford, OX1\,3PU, United Kingdom}
\author{Yannik Kunz}
\affiliation{Fachbereich Physik and Landesforschungszentrum OPTIMAS, Rheinland-Pfälzische Technische Universität Kaiserslautern-Landau, 67663 Kaiserslautern, Germany}
\author{Timmy Reimann}
\affiliation{INNOVENT e.V. Technologieentwicklung, 07745 Jena, Germany}
\author{Morris Lindner}
\affiliation{INNOVENT e.V. Technologieentwicklung, 07745 Jena, Germany}
\author{Carsten Dubs}
\affiliation{INNOVENT e.V. Technologieentwicklung, 07745 Jena, Germany}
\author{John F. Gregg}
\affiliation{Clarendon Laboratory, Department of Physics, University of Oxford, Parks Road, Oxford, OX1\,3PU, United Kingdom}
\author{Mathias Weiler}
\affiliation{Fachbereich Physik and Landesforschungszentrum OPTIMAS, Rheinland-Pfälzische Technische Universität Kaiserslautern-Landau, 67663 Kaiserslautern, Germany}

\date{\today}%

\begin{abstract}
We study surface acoustic waves (SAWs) in yttrium iron garnet (YIG)/zinc oxide (ZnO) heterostructures, comparing the results of a computationally lightweight analytical model with time-resolved micro-focused Brillouin light scattering ($\mu$-BLS) data.  Interdigital transducers (IDTs), with operational frequencies in the gigahertz regime, were fabricated on 50 and 100\,nm thin films of YIG prior to sputter deposition of 830\,nm and 890\,nm films of piezoelectric ZnO. We find good agreement between our analytical model and $\mu$-BLS data of the IDT frequency response and SAW group velocity, with clear differentiation between the Rayleigh and Sezawa-like modes. This work paves the way for the study of SAW-spin wave (SW) interactions in low SW damping YIG, with the possibility of a method for future energy-efficient SW excitation. 
\end{abstract}

\maketitle

\section{Introduction}

Surface acoustic waves (SAWs) have become ubiquitous in modern life, indeed most of us carry SAW devices every day in the form of mobile telephone filters~\cite{campbell1998surface, 7153041}, owing to their relatively short wavelengths at gigahertz frequencies compared to their electromagnetic counterparts. Other applications include sensors~\cite{gronewold2007surface, mandal2022surface}, oscillators~\cite{wohltjen1984mechanism, parker1988precision}, and microfluidic actuators~\cite{ding2013surface, destgeer2015recent}; however, of late, there has been increasing research on the coupling between SAWs and spin waves (SWs) in thin magnetic films~\cite{sasaki2017nonreciprocal, tateno2020highly, hernandez2020large, 
 xu2020nonreciprocal, kuss2020nonreciprocal, kuss2021symmetry, kuss2021nonreciprocal, li2021advances, geilen2022fully, kuss2023nonreciprocal1, kuss2023nonreciprocal, huang2023phonon, kunz2023coherent}. This has led to the observation of several intriguing phenomena, such as non-reciprocal SW generation/ SAW absorption as a result of a mismatched helicity between the SAW-induced magnetoacoustic driving fields and the fixed precession of the magnetisation~\cite{sasaki2017nonreciprocal, xu2020nonreciprocal, kuss2021symmetry}, with possible application as acoustic isolators or circulators~\cite{shah2020giant, verba2021phase, rasmussen2021acoustic, kuss2022chiral}. Furthermore, there is the prospect of utilising SAWs for the energy-efficient excitation of SWs, due to the absence of Joule heating compared to conventional microwave antenna or spin pumping via the spin-hall effect, with application in magnonic computing~\cite{li2017spin, mahmoud2020introduction, chumak2022advances}. However, these SAW-SW studies have suffered from high SW damping in the magnetic materials of interest. For example, CoFeB has propagation lengths typically on the order of micrometres~\cite{rana2017excitation, nikitchenko2021spin}. 

In this work, piezoelectric zinc oxide (ZnO) was deposited by radio frequency (RF) magnetron sputtering on thin films of yttrium iron garnet (YIG), which exhibits SW propagation lengths of up to millimetres~\cite{qin2018propagating, maendl2017spin}. Zinc oxide (ZnO) is a well-established piezoelectric material, capable of generating high-frequency SAWs on account of its relatively high acoustic wave velocity and electromechanical coupling coefficient~\cite{koike19931, le2008ghz, wang2008gigahertz, fu2018high, su2020enhanced}. Although YIG/ZnO heterostructures have been realised in the past ~\cite{hanna1988interactions, kryshtal2017nonlinear}, the interdigital transducers (IDTs) fabricated were not capable of generating SAWs in the gigahertz regime, which is required for the coupling of surface acoustic waves and spin waves. Before sputter deposition, we use electron-beam lithography to pattern IDTs on the YIG with a periodicity of 2.8\,$\mu$m and a corresponding fundamental frequency of 1.1\,GHz. The IDTs can also be operated at higher harmonics of the fundamental frequency, where the first accessible harmonic frequency for our IDT design is 2.9\,GHz. This is noteworthy, as with the harmonic IDT excitation we can reach sufficiently high frequencies to enable the direct study of the SAW-SW interaction in YIG in the future.

In this paper, we report on the SAW properties in these YIG/ZnO heterostructures. We measure the frequency response of the IDTs and the group velocities of the launched SAWs, using time-resolved micro-focused Brillouin light scattering ($\mu$-BLS), to characterise these devices. Furthermore, we compare these results to a computationally lightweight analytical model that calculates the SAW dispersion relation in the heterostructures, taking the entire stack sequence substrate/YIG/ZnO into account. We find that this model is in good agreement with our experimental results.

\section{Theory \label{sec:theory}}
\begin{figure}[!ht]
\centering
\vspace{0pt}
\includegraphics[width=0.48\textwidth]{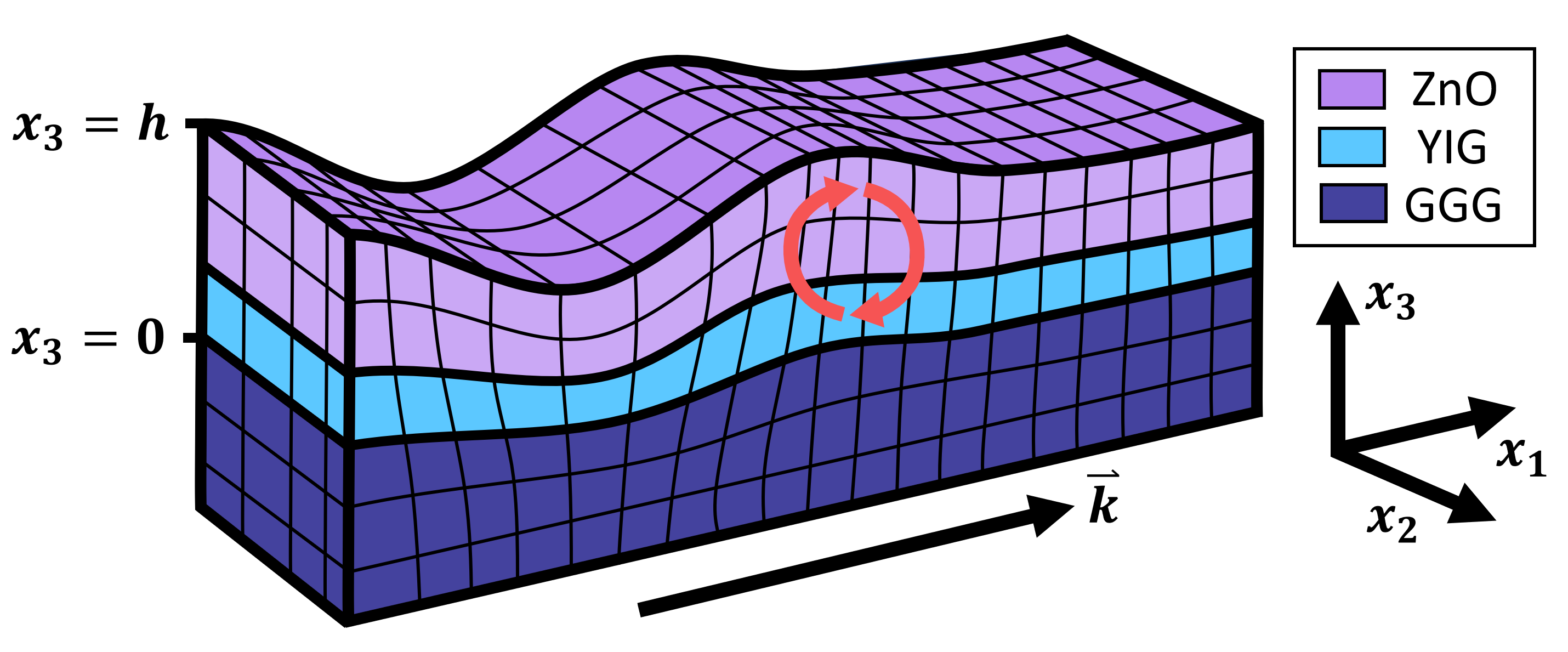}
\caption{Schematic of the geometry of the thin film structures used in the analytical model. Thin layers of YIG and ZnO on a GGG substrate are shown. The direction of wave propagation and the coordinate system are indicated by $\boldsymbol{k}$, $\boldsymbol{x_1}$, $\boldsymbol{x_2}$, $\boldsymbol{x_3}$. A Rayleigh wave is shown propagating in the $\boldsymbol{x_1}$ direction, where the red arrows indicate the characteristic elliptical motion of the lattice in the $\boldsymbol{x_1}$, $\boldsymbol{x_3}$ plane.}
\label{fig:geometry}
\vspace{0pt}
\end{figure}

To begin, we present the analytical model used to determine the SAW dispersion relation in a thin-film structure consisting of $n$-layers. The methodology follows that laid out by Farnell and Adler~\cite{FARNELL197235}; however, we consider it valuable to present it here in some detail to aid in the comparison between the model and our experimental results. We consider the model a useful tool to obtain fairly accurate analytical results in just a few minutes, without the need to resort to computationally intensive finite-element modeling. The relevant geometry of the thin-film layered structures can be seen in Fig.~\ref{fig:geometry}, where we show a YIG/ZnO two-layer thin film structure on a Gadolinium Gallium Garnet (GGG) substrate. $\boldsymbol{x_3}$ is the direction normal to the surface, with the interface between the first layer and the substrate at $x_3 = 0$, and a total YIG/ZnO film layer thickness given by $x_3 = h$. The GGG substrate is assumed to extend to $-\infty$ in the $\boldsymbol{x_3}$ direction. The waves propagate in the $\boldsymbol{x_1}$ direction.

For a piezoelectric material, such as ZnO, the equations of motion are given by
\begin{equation}
\begin{split}
    \rho \pdv[2]{u_j}{t} = c_{ijkl} \pdv[2]{u_k}{x_i}{x_l} + e_{kij} \pdv[2]{\phi}{x_i}{x_k} \\ 
    e_{ikl} \pdv[2]{u_k}{x_i}{x_l} = \epsilon_{ik} \pdv[2]{\phi}{x_i}{x_k},
\label{eq:EoM}
\end{split}
\end{equation}
where $u_j$ are the mechanical displacements, $\phi$ the electric potential, $\rho$ the density, and $c_{ijkl}$, $e_{kij}$ and $\epsilon_{ik}$ the elastic, piezoelectric, and permittivity tensors respectively.

As we are looking for SAWs, we propose the ansatz
\begin{equation}
\begin{split}
    u_j & = a_je^{ikbx_3}e^{ik(x_1-vt)}\\
    \phi & = a_4e^{ikbx_3}e^{ik(x_1-vt)},
\label{eq:ansatz}
\end{split}
\end{equation}
where we have a wave of amplitude $a_j$ propagating in the $x_1$ direction with wavenumber $k$ and phase velocity $v$. The exponentially decaying component in the $x_3$ direction, with complex coefficient $b$, gives the SAW characteristic. Substituting this ansatz into the equations of motion yields the Christoffel equation
\begin{equation}
\label{eq:Christoffel}
\begin{aligned}
\begin{pmatrix}
\Gamma _{11} -\rho v^{2} & \Gamma _{12}             & \Gamma _{13}             & \Gamma _{14}\\
\Gamma _{12}             & \Gamma _{22} -\rho v^{2} & \Gamma _{23}             & \Gamma _{24}\\
\Gamma _{13}             & \Gamma _{23}             & \Gamma _{33} -\rho v^{2} & \Gamma _{34}\\
\Gamma _{14}             & \Gamma _{24}             & \Gamma _{34}             & \Gamma _{44}
\end{pmatrix}\begin{pmatrix}
a_{1}\\
a_{2}\\
a_{3}\\
a_{4}
\end{pmatrix}\\
= 0.
\end{aligned}
\end{equation}
$\Gamma_{ij}$ are given by quadratic equations in $b$ with components of the elastic, piezoelectric, and permittivity tensors as the coefficients. 

For a nontrivial solution, the determinant of the matrix in eq.\,(\ref{eq:Christoffel}) must be zero; therefore, for each value of $v$, there is an eighth-order polynomial in $b$ to solve. The amplitude coefficients, $a_j$, can then be found by solving eq.\,(\ref{eq:Christoffel}) for each solution of $b$; giving a solution that is a superposition of partial waves
\begin{equation}
\begin{split}
u_j =\sum_{m} C_{m} a_{j}^{(m)}e^{ikb^{(m)} x_{3}} e^{ik( x_{1} -vt)}\\
\phi =\sum_{m} C_{m} a_{4}^{(m)}e^{ikb^{(m)}x_{3}} e^{ik( x_{1} -vt)},
\label{eq:partialwaves}
\end{split}
\end{equation}
rather than the monochromatic wave ansatz in eq.\,(\ref{eq:ansatz}). 

As we are considering thin layers, where the wavelengths are comparable to the thicknesses, all the solutions of $b^{(m)}$ are taken, such that the index $m$ runs from 1 to 8. However, in the substrate, only the $b^{(m)}$ in the lower half of the complex plane are considered, as the wave must vanish as $x_3\rightarrow -\infty$, hence the index $m$ runs from 1 to 4. Therefore, in total, the index $m$ runs from 1 to 8$\times$(number of layers)+4.

The solutions found thus far only give the relations between $b^{(m)}$, $a_i^{(m)}$, and $v$. As we are interested in the dispersion relations, we must additionally consider the boundary conditions. To do so, we use the linearised coupled strain-charge equations
\begin{equation}
\begin{split}
T_{ij} = c_{ijkl}S_{kl} - e_{kij}E_{k}\\
D_{i} = e_{ikl}S_{kl}+\epsilon_{ik}E_{k},
\label{eq:linchargecoupled}
\end{split}
\end{equation}
where $T_{ij}$ is the stress tensor, $S_{kl}$ the strain tensor, $E_{k}$ the electric field, and $D_{k}$ the electric displacement with
\begin{equation}
S_{kl} = \frac{1}{2}\left( \pdv{u_k}{x_l}  + \pdv{u_l}{x_k} \right) \text{and} \,\, E_k = - \pdv{\phi}{x_k}. 
\label{eq:strain}
\end{equation}

At each layer interface, $u_j$, $\phi$, $D_3$, $T_{23}$, $T_{13}$, and $T_{33}$ are continuous, giving 8 boundary condition equations. At the free surface, there is no restriction on the mechanical displacements, while the same three stress components are zero, and $D_3$ is again continuous; giving four boundary condition equations. Therefore, in total, there are 8$\times$(number of layers)+4 boundary condition equations.

Substituting eq.\,(\ref{eq:partialwaves}) into the boundary condition equations, gives linear equations in the relative amplitudes $C_m$, and hence can be written in the form of a matrix equation
\begin{equation}
BC_{pm} C_m = 0,
\label{eq:BC}
\end{equation}
where $BC_{pm}$ is the boundary condition matrix and the index $p$ runs over the boundary condition equations. As before, for nontrivial solutions, the determinant of $BC_{pm}$ must be zero. Solving this equation gives the phase velocity of the structures in terms of the wavenumber, from which the dispersion relation and group velocity can be easily determined. 

In addition, the electromechanical coupling coefficients, given by~\cite{zhang_2022_bulk}
\begin{equation}
K^2 = 2 \left( 1 - \frac{ v_{\mathrm{metalised}} }{v_{\mathrm{free}}} \right),
\label{eq:couplingcoefficient}
\end{equation}
can be calculated. Where $v_{\mathrm{free}}$ is the phase velocity previously calculated, and $v_{\mathrm{metalised}}$ is the phase velocity calculated with an infinitesimally thin perfect conductor layer at the position of the IDT. This metallic film modifies the boundary conditions such that the electric potential, $\phi$, at the conductor is zero. If the film is on the free surface, the size of the boundary condition matrix is unchanged; whereas, in the case of an interlayer IDT, an additional 8 equations must be added. $K^2$ gives an estimate of the conversion efficiency between electrical and acoustic energy.

Finally, by solving eq.\,(\ref{eq:BC}), the relative amplitudes $C_m$ can be calculated. With all the coefficients calculated, eq.\,(\ref{eq:partialwaves}) can be solved to find the normalised mechanical displacements as a function of depth, and the strains calculated using eq.\,(\ref{eq:strain}).

\section{Sample structuring}

\begin{figure}[!ht]
\centering
\vspace{0pt}
\includegraphics[width=0.48\textwidth]{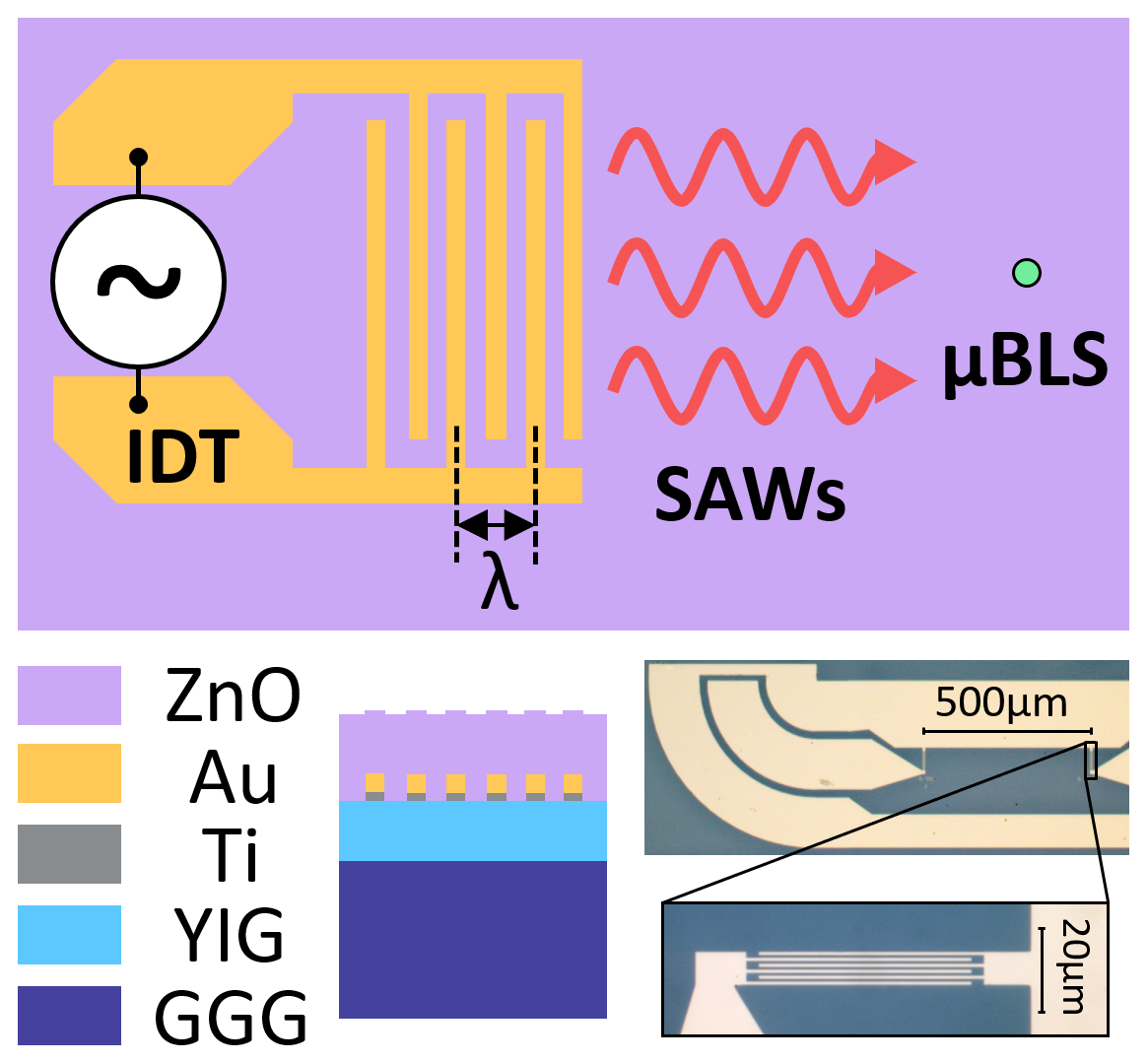}
\caption{Top: schematic of the $\mu$-BLS experiment showing an interdigital transducer (IDT) excited by a microwave source, the induced surface acoustic waves are indicated, as is the laser spot used to measure the phonons. Bottom left: schematic of the sample structure showing the layers and the embedded IDTs. Bottom right: microscope image of one of the samples, a gold IDT structure can be seen as the lighter colour with the IDT fingers highlighted by the inset.}
\label{fig:devicestructure}
\vspace{0pt}
\end{figure}

Before examining the results of the analytical model, we first discuss the sample preparation and structuring. A schematic of the sample structure and a microscope image of one of the IDTs can be seen in Fig.~\ref{fig:devicestructure}. Experimental results from two samples are presented here and the thicknesses of the layers and IDTs can be found in Tab.~\ref{table:Samples}. 

\begin{table}[ht]
\caption{The thicknesses of the material layers and IDTs of the two samples.} % title of Table
\centering % used for centering table
\begin{tabular}{c c c c c c} % centered columns (4 columns)
\hline\hline %inserts double horizontal lines
\multirow{2}{*}{Sample} & \multirow{2}{*}{GGG} & \multirow{2}{*}{YIG} & \multicolumn{2}{c}{IDT} & \multirow{2}{*}{ZnO} \\
& & &Ti& Au& \\[0.5ex] 
\hline % inserts single horizontal line
Sample 1 & 0.5mm & 103nm & 10nm & 80nm & 890nm\\ 
Sample 2 & 0.5mm & 55nm & 10nm & 80nm & 830nm\\[0.5ex] % 
\hline %inserts single line
\end{tabular}
\label{table:Samples} % is used to refer this table in the text
\end{table}

The YIG thin films were grown by liquid phase epitaxy on a (111) oriented GGG substrate~\cite{dubs2020low}. The 103\,nm film was purchased commercially from the company Matesy GmbH in Jena, Germany, and the 55\,nm film was grown at INNOVENT e.V., Germany. To reach sufficiently high frequencies to enable the future study of SAW-SW coupling, electron beam lithography was used to pattern IDTs with 6 and 20-fingers with finger widths and separations of 700\,nm ($\lambda =$\,2.8\,$\mu$m), and an aperture size of 50\,$\mu$m. The titanium-gold stack was deposited by electron beam evaporation, followed by a lift-off process to leave behind the IDT. The ZnO was RF magnetron sputtered over the entire sample, with the piezoelectric c-axis of the wurtzite crystal structure pointing out-of-plane. Finally, the contact pads of the IDTs were etched free of the insulating ZnO using hydrochloric acid. A more detailed description of this process, including x-ray diffraction (XRD) characterisation, can be found in the Supplemental Material~\cite{SupplemetalMaterial}.

\section{Experimental methods}

To excite SAWs, the IDT is contacted using a microwave probe with a ground-signal-ground footprint and pitch of 200\,$\mu$m, connected to a microwave source and an amplifier. The IDT is directly excited and the SAW intensity is measured using $\mu$-BLS~\cite{TimetaggerPaper, demidov2004radiation, TimetaggerFigure}, with a 532\,nm wavelength laser, to determine the IDT's frequency response. A schematic of this set-up can be seen in Fig.~\ref{fig:devicestructure}. Note, a half-wave plate is used to suppress possible magnon-induced signals by their polarization-dependence~\cite{kunz2023coherent}. 

\begin{figure}[!ht]
\centering
\vspace{0pt}
\includegraphics[width=0.48\textwidth]{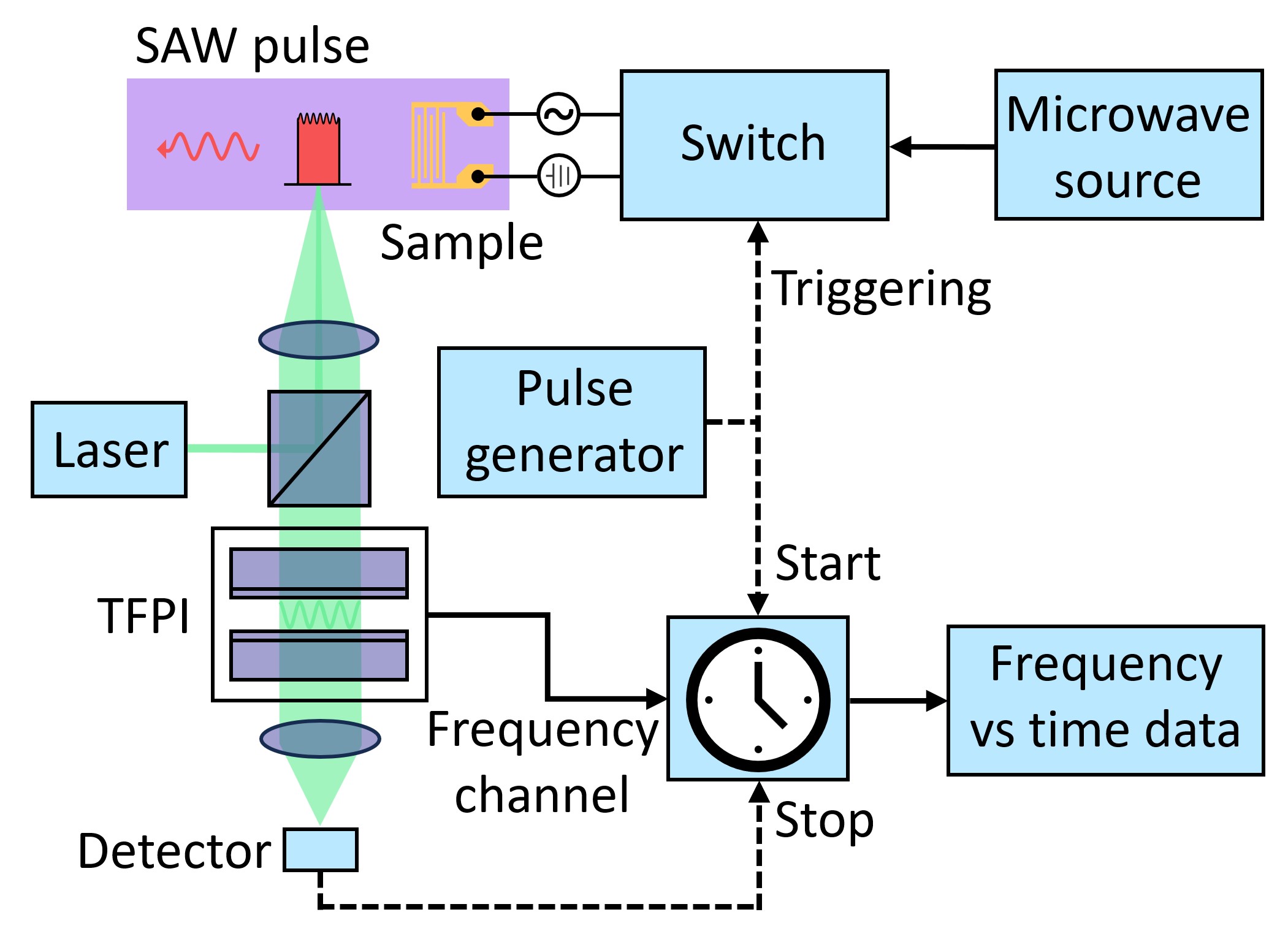}
\caption{Schematic of the time-resolved micro-focused Brillouin light scattering (BLS) spectroscopy setup. A pulse generator is used to trigger the start and end of a BLS measurement window during which a SAW pulse, well defined in time, is also triggered. This pulse is realised using a microwave source and a fast microwave switch. Figure adapted with permission from~\cite{heinz2021nano}.}
\label{fig:timetagger}
\vspace{0pt}
\end{figure}

In addition to standard $\mu$-BLS measurements, we also take time-resolved data to find the phonon group velocities. A schematic of the time-resolved $\mu$-BLS setup is shown in Fig.~\ref{fig:timetagger}. A pulse generator is used to trigger a $\mu$-BLS measurement window, during this window a microwave switch is opened for approximately 600\,ns, allowing a microwave pulse of fixed frequency to excite the IDT. 

\section{Analytical results \label{sec:analytical}}

Considering the above sample structures, we calculated the dispersion relations following the methodology laid out in Section~\ref{sec:theory}. This requires a large number of material parameters, as discussed below, and there are no free parameters in the calculation, as such, an exact solution is numerically converged upon. Practically, we propose an initial guess for $k$ at a fixed $v$, and if a solution for $k$ is converged upon, continue to make small steps in $v$ over a predefined range finding $k(v)$. 

\begin{figure*}[!ht]
\subfloat[\label{fig:Dispersion}]{%
  \includegraphics[width=.49\textwidth]{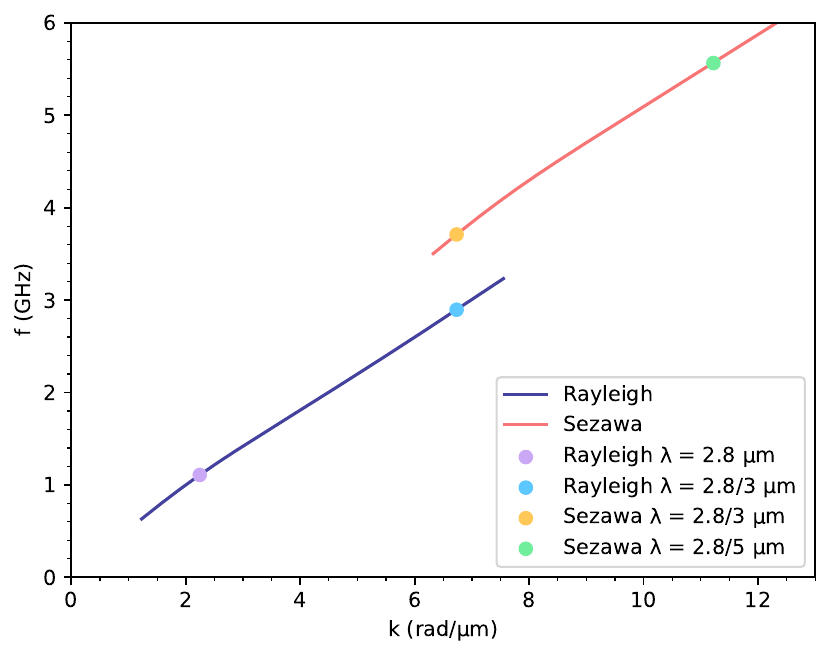}%
}\hfill
\subfloat[\label{fig:FreqResponse}]{%
  \includegraphics[width=.49\textwidth]{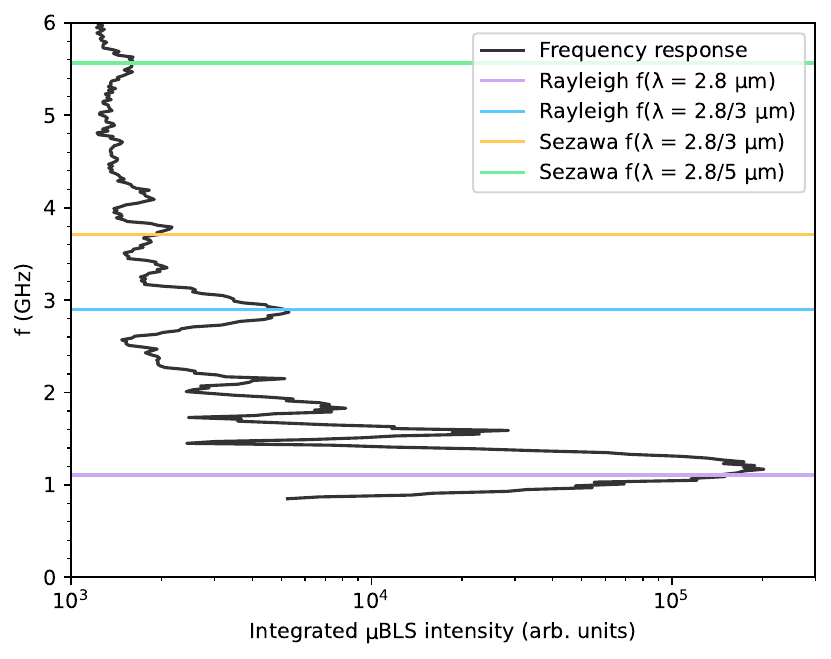}%
}
\caption{(a) The calculated dispersion relation for the Rayleigh-like mode, dark blue line, and the first Sezawa-like mode, red line, for Sample 1. The four points indicate the expected excitation frequencies of the fixed $\lambda =$\,2.8\,$\mu$m transducers. (b) Experimental data, in dark grey, showing the frequency response of the 6-finger interdigital transducer on Sample 1, as measured by micro-focused Brillouin light scattering. The coloured horizontal lines indicate the resonant frequencies at which the $\lambda =$\,2.8\,$\mu$m transducer is expected to excite surface acoustic waves as calculated by the analytical model.}
\end{figure*}

The resulting dispersion relation for Sample 1 can be seen in Fig.~\ref{fig:Dispersion}. Solving the wave equation leads to multiple solutions, which can be grouped into two classes depending on the type of particle displacement induced. The first are `Rayleigh modes', which result in an elliptical particle motion in the $\boldsymbol{x_1}$, $\boldsymbol{x_3}$ plane for the case of isotropic materials~\cite{rayleigh1885waves, FARNELL197235}, as shown schematically in Fig.~\ref{fig:geometry} by the red arrows. We note that due to the anisotropic layered structures, there is additional particle motion as discussed in the Supplemental Material~\cite{SupplemetalMaterial}, hence we use the term `Rayleigh-like'. The zeroth-order mode, shown by the dark blue line in Fig.~\ref{fig:Dispersion}, always exists in the structure; however, the higher-order `Sezawa' modes are only introduced when the shear wave speed in the layer exceeds that of the substrate~\cite{Sezawa1927DispersionOE}. The first-order Sezawa-like mode is shown by the red line in Fig.~\ref{fig:Dispersion}. 

The second class of solutions are `Love modes', which have particle motion perpendicular to the direction of propagation for isotropic materials~\cite{love1911some}. In the case of our layered structures, we find the expected excitation frequencies of the Love-like modes to be similar to those of the Rayleigh-like modes, thus prohibiting their differentiation with BLS frequency response data due to the finite linewidths. However, the difference between the Rayleigh and Love-like modes is significantly more pronounced when considering the group velocity curves. We did not find the calculated group velocity curves for the Love-like modes to fit well with the experimental data, and therefore, chose to ignore this class of solutions. This is presumably owing to the small electromechanical coupling coefficients calculated for these modes, typically two orders of magnitude smaller than those of the Rayleigh-like modes, another reason we chose to ignore these solutions.

It is important to note a few caveats relating to our methodology: firstly, the IDTs, in particular their finite thickness, are neglected in the simple analytical model; secondly, as mentioned, literature values for the density and elastic, piezoelectric, and permittivity tensors are used for the ZnO~\cite{zhang_2022_bulk}, YIG~\cite{clark1961elastic, hasan2017yttrium}, and GGG~\cite{graham1970elastic, connelly2021complex} as listed in the Supplemental Material~\cite{SupplemetalMaterial}, rather than measured values for our films. As this results in a total of 23 material parameters used in the model, generally measured in bulk samples, we do not attempt to estimate an error in the analytical calculation. 

Thirdly, based on our experience of growing ZnO films sputtered under near identical conditions, we expect there to be an approximately 200\,nm `dead layer' of ZnO where the c-axis is not aligned~\cite{fung2015phonon}. This layer occurs as the ZnO does not begin growth with an aligned c-axis, instead growing with a random orientation that begins to self-texture, preferentially forming out-of-plane c-axis aligned columnar grains. To account for this, we introduce an additional 200\,nm layer between the YIG and the c-axis aligned ZnO, which we treat as polycrystalline ZnO with randomly oriented c-axes - calculating the material constants accordingly~\cite{den1999relation}. Physically, we expect a gradual increase in c-axis alignment from the initial complete randomisation; however, we consider the simple model to be adequate as a first-order approximation given the observed change in the dispersion relation is minor. The only sizeable effect is a reduction in the electromechanical coupling coefficients by one to two orders of magnitude, depending on frequency. This is expected, given the electrical excitation from the embedded IDTs is concentrated around this `dead layer' where there is no macroscopic alignment of electric dipoles. Indeed, improving c-axis alignment and reducing the `dead layer' thickness is the key to improved excitation efficiency and an area of active research. We consider these three caveats to be the most likely source of any discrepancies between analytical and experimental results; although, as will be shown, these are relatively small.  

Further discussion of the model of the `dead layer', the rotation of crystallographic axes including the ZnO piezoelectric c-axis, and calculations of the coupling coefficients, and normalised mechanical displacements and strains may be found in the Supplemental Material~\cite{SupplemetalMaterial}.

\section{Experimental Results}

First, we measured the frequency response of the 6-finger IDT on Sample 1 and compared this to the expected excitation frequencies from the analytical calculation. This comparison is shown in Fig.~\ref{fig:FreqResponse}. To maximise the signal, the laser spot was positioned centrally relative to the IDT aperture and approximately 1\,$\mu$m along the SAW propagation path. We expect the highest intensity peak to correspond to the fundamental frequency which is determined by the periodicity/ wavelength of the IDT (2.8\,$\mu$m) according to $f=v/\lambda$. Moreover, we should see higher order modes corresponding to $\lambda/p$ where $p$ is an odd integer - a net electric field between IDT fingers is required to excite a SAW. We see the fundamental frequency occurs at approximately 1.1\,GHz. The coloured horizontal lines indicate the calculated excitation frequencies of the IDT and show good agreement with the peaks in the experimental data.

\begin{figure*}[!ht]
\subfloat[\label{fig:Linescans}]{
  \includegraphics[width=.49\textwidth]{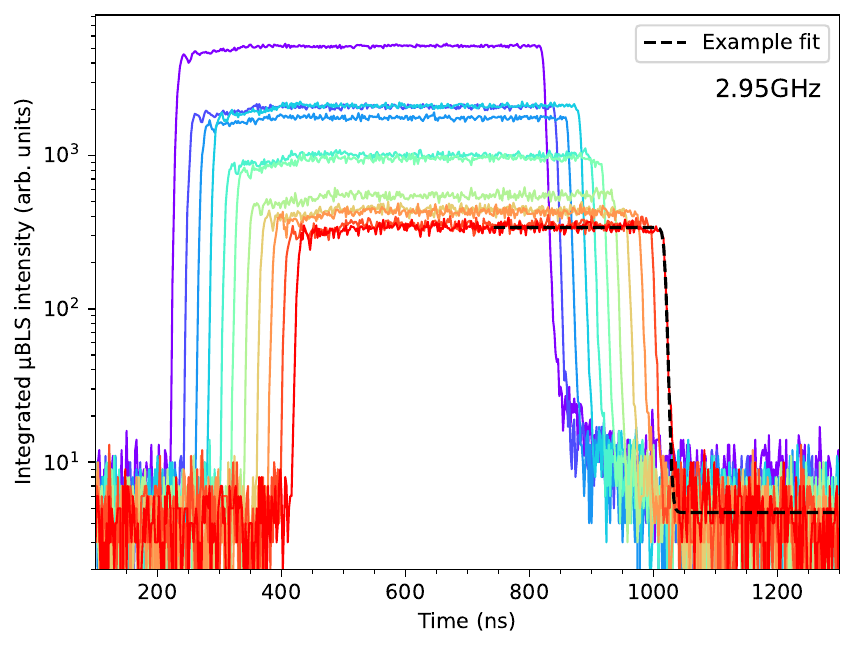}%
}\hfill
\subfloat[\label{fig:VelocityFit}]{%
\includegraphics[width=.49\textwidth]{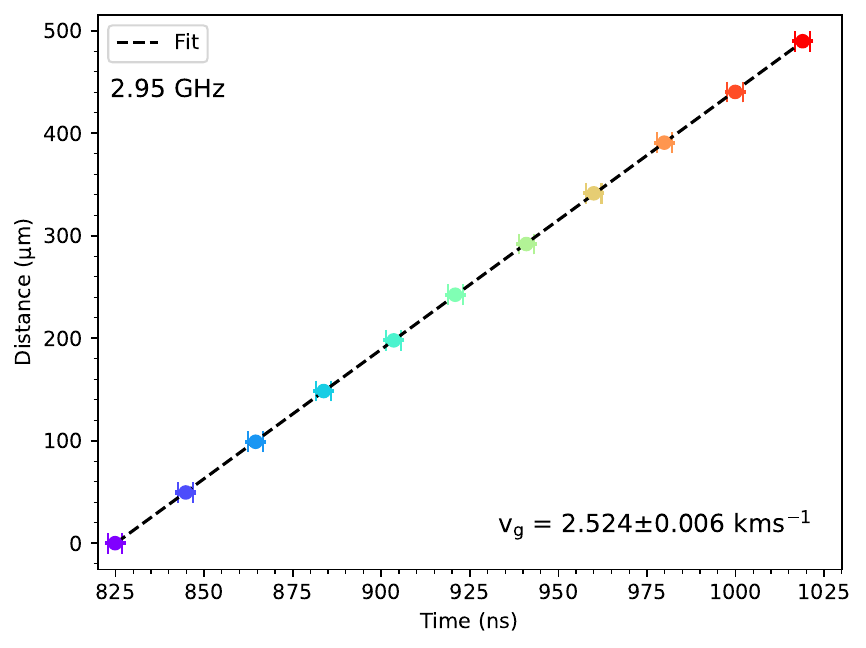}%
}\hfill
\subfloat[\label{fig:Sample11_GroupVelocity}]{%
  \includegraphics[width=.49\textwidth]{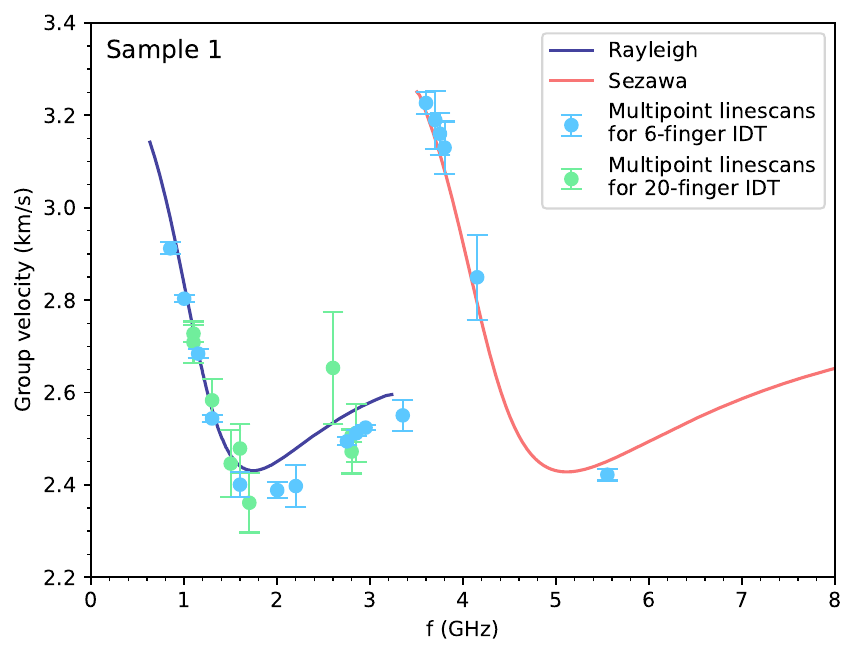}%
}\hfill
\subfloat[\label{fig:Sample19_GroupVelocity}]{%
  \includegraphics[width=.49\textwidth]{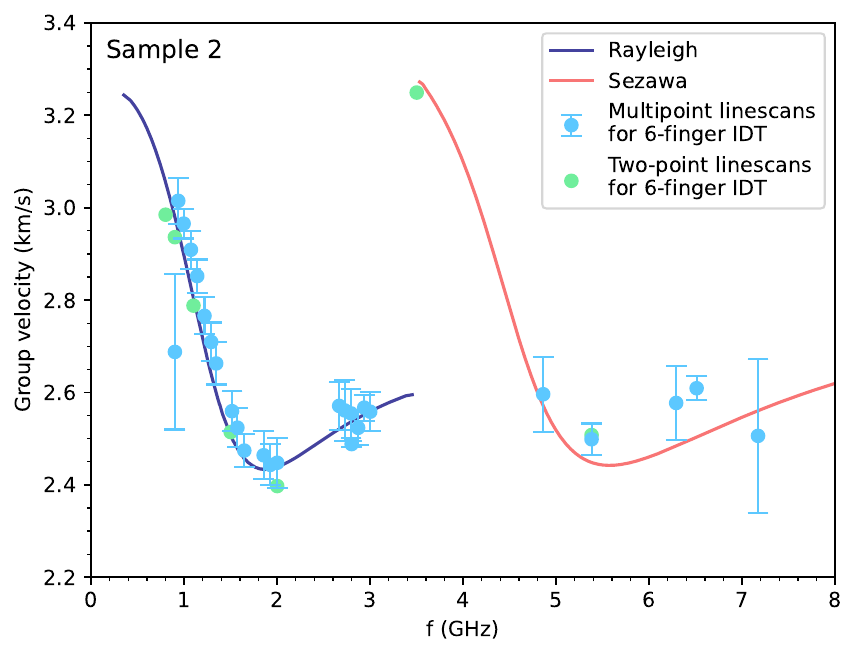}%
}
\caption{(a) Experimental data showing measured BLS intensity as a function of time. The different colours show BLS data taken at different laser spot positions, purple corresponds to the position closest to the IDT and the colour progression through to red goes with increasing SAW propagation distance. The black dotted line shows a Boltzmann function fitted to the red data taken furthest from the IDT. (b) The time at the centre of the falling edge of the SAW pulse, extracted by fitting the data in Fig.~\ref{fig:Linescans}, as a function of the distance of the laser spot from the IDT. The dotted black line was fitted to find the group velocity. All the data for (a) and (b) was taken at a fixed microwave excitation frequency of 2.95\,GHz on Sample 1. The calculated Rayleigh and Sezawa-like mode group velocities, as a function of frequency, are shown by the dark blue and red lines in (c) and (d). (c) Experimental group velocity data for Sample 1, the light blue points show data for a 6-finger IDT and the green points for a 20-finger IDT. (d) Experimental group velocity data for a 6-finger IDT on Sample 2. In blue, multipoint linescans, where the laser spot is scanned over multiple points in space, are shown. In green, two-point linescans, where the laser spot measures at only two points in space, are shown.}
\end{figure*}

Second, we used time-resolved measurements to calculate the phonon group velocities at fixed microwave frequencies. An example of the resultant BLS intensity as a function of time can be seen in Fig.~\ref{fig:Linescans}. We see top-hat-like intensity profiles, where high intensities correspond to the presence of a SAW at the laser spot position. The laser spot is initially positioned centrally next to the IDT and then moved away from the IDT along the SAW propagation direction in equal-sized steps. At each step, a BLS measurement is taken corresponding to the different coloured lines from purple to red in Fig.~\ref{fig:Linescans}. This data is then fitted using a least squares regression with the sigmoidal Boltzmann function to smoothly approximate the Heaviside step function
\begin{equation}
    I = C + \frac{A-C}{1+e^{-\frac{t-t_0}{B}}},
\end{equation}
where $I$ is the BLS intensity, $t$ is the time, and $A$, $B$, $C$, and $t_0$ are constants. An example fit is shown by the black dotted curve fitting the red BLS data.

From these fits we extract the constant $t_0$, the position of the falling edge of the SAW pulse, with an associated fitting error. Combining this with the known laser spot position, we can plot the SAW propagation distance from the IDT as a function of $t_0$, Fig.~\ref{fig:VelocityFit}. Alongside the fitting error in $t_0$, we take into account experimental errors of 2\,ns for time-based measurements and 1\,$\mu$m for the microscope position stabilisation. A straight line is fitted to the data, using orthogonal distance regression, to determine the group velocity at a fixed frequency.

The results of this fitting process can be seen for Sample 1 in Fig.~\ref{fig:Sample11_GroupVelocity} and Sample 2 in Fig.~\ref{fig:Sample19_GroupVelocity}. We show the experimentally determined group velocities with their associated errors in light blue and green as a function of the SAW excitation frequency. For Sample 1, the group velocities were measured for two IDT structures, one with 6-fingers, light blue, and the other with 20-fingers, green. These group velocities were determined from multipoint linescans, that is to say, the laser spot was scanned over multiple points in space, as in Fig.~\ref{fig:Linescans}. For Sample 2, all measurements were made on the same 6-finger IDT. In light blue, we again have group velocities determined from multipoint linescans. In green, however, the group velocities were calculated from two-point linescans, where the laser spot was positioned at only two points in space, one next to the IDT and the other at the maximum distance measured from the IDT of 500\,$\mu$m. There is good agreement between the two-point and multipoint measurements showing the accuracy of the technique. 

Figs~\ref{fig:Sample11_GroupVelocity} and~\ref{fig:Sample19_GroupVelocity} also show the analytically calculated group velocities of the Rayleigh-like mode in dark blue and the Sezawa-like mode in red for Sample 1 (Fig~\ref{fig:Sample11_GroupVelocity}) and Sample 2 (Fig~\ref{fig:Sample19_GroupVelocity}). Given the caveats discussed in Section~\ref{sec:analytical}, the agreement between the analytical model and the experimental data is excellent, with the largest discrepancies, in general, occurring for the points with the largest measurement errors. These large errors occur due to the low excitation efficiency of the IDT at these frequencies, meaning the top-hat-like intensity profiles become more noisy. In particular, the narrower bandwidth 20-finger IDT tends to show larger errors when off-resonance, as expected. In general, the data fits the curvature well and we can differentiate between the Rayleigh and Sezawa-like modes. These results show we can measure the non-linearity in the phonon dispersion relation for these complex layered structures to a high accuracy using time-resolved $\mu$-BLS. Furthermore, we demonstrate good agreement with the analytical model, thus verifying the model is sufficient within the assumptions made to interpret our experimental data. Additionally, the experimental results agree with the model for two different samples and three different IDT structures with different ZnO and YIG thicknesses.

Finally, from the amplitudes of the top-hat-like functions found from the Boltzmann fits, we can estimate the phonon decay length by fitting the equation
\begin{equation}
    A(x_1) = A_0 e^{-\frac{x_1}{\Lambda}},
\end{equation}
where $A_0$ is the initial amplitude and $\Lambda$ the decay length. Due to the aforementioned reduction in experimental data quality when off-resonance, we only calculate this decay length for values near the first accessible harmonic frequency of the IDT. Taking a weighted average, we find a decay length of $127\pm31\,\mu\text{m}$ for the 6-finger IDT on Sample 1, consistent with the value of $124\pm65\,\mu\text{m}$ for the 6-finger IDT on Sample 2. We believe the relatively large errors in these values result from variations in the reflectivity of the surface of the samples, which could result from defects in the film or surface particles. We note that etching away the ZnO along the propagation path after the excitation region may increase the decay length, given the `dead layer' is not crystallographically ordered, and therefore, will have increased scattering.

\section{Conclusion and outlook}

In conclusion, we have demonstrated a method to generate gigahertz frequency SAWs in YIG thin films by first patterning IDTs on the surface of the YIG, and then sputter depositing thin films of piezoelectric ZnO. We measured the IDT frequency responses and SAW group velocities of these structures experimentally using time-resolved $\mu$-BLS. Using these experimental results, we show good agreement with an analytical model that calculates the dispersion relation of the SAWs in these thin film heterostructures. We are able to differentiate between Rayleigh and Sezawa-like modes. 

We hope this work lays the foundations for the study of the SAW-SW interaction in YIG, with the possibility of observing non-reciprocal SW generation, and therefore, probing device ideas such as acoustic isolators or circulators~\cite{sasaki2017nonreciprocal, xu2020nonreciprocal, kuss2021symmetry}. With efficient SAW excitation, there is additionally the prospect of studying strong magnon-phonon coupling~\cite{hwang2023strongly} and non-linear SAW-SW interaction phenomena~\cite{geilen2022parametric, shah2023symmetry}. Furthermore, the successful deposition of piezoelectric ZnO, despite the large lattice mismatch with YIG~\cite{Abrahams1969Remeasurement, Gurjar2019Structural}, suggests it may be possible to similarly excite SAWs in other lattice mismatched thin film materials. This demonstrates the versatility of using ZnO as a piezoelectric and possibly opens the door to studying the magnetoacoustic interaction in additional magnetic media. 

\begin{acknowledgments}
This work was supported by the European Research Council (ERC) under the European Union’s Horizon Europe research and innovation programme (Consolidator Grant ``MAWiCS", Grant Agreement No. 101044526). The work of F. Ryburn was supported by a UK Engineering and Physical Sciences Research Council (EPSRC) Industrial Cooperative Award in Science \& Technology. The work of M. Linder was supported by the German Bundesministerium f{\"u}r Wirtschaft und Energie 262 (BMWi) under Grant No. 49MF180119.
\end{acknowledgments}

\bibliographystyle{apsrev4-1}
\bibliography{Bibliography}

\begin{thebibliography}{69}%
\makeatletter
\providecommand \@ifxundefined [1]{%
 \@ifx{#1\undefined}
}%
\providecommand \@ifnum [1]{%
 \ifnum #1\expandafter \@firstoftwo
 \else \expandafter \@secondoftwo
 \fi
}%
\providecommand \@ifx [1]{%
 \ifx #1\expandafter \@firstoftwo
 \else \expandafter \@secondoftwo
 \fi
}%
\providecommand \natexlab [1]{#1}%
\providecommand \enquote  [1]{``#1''}%
\providecommand \bibnamefont  [1]{#1}%
\providecommand \bibfnamefont [1]{#1}%
\providecommand \citenamefont [1]{#1}%
\providecommand \href@noop [0]{\@secondoftwo}%
\providecommand \href [0]{\begingroup \@sanitize@url \@href}%
\providecommand \@href[1]{\@@startlink{#1}\@@href}%
\providecommand \@@href[1]{\endgroup#1\@@endlink}%
\providecommand \@sanitize@url [0]{\catcode `\\12\catcode `\$12\catcode `\&12\catcode `\#12\catcode `\^12\catcode `\_12\catcode `\%12\relax}%
\providecommand \@@startlink[1]{}%
\providecommand \@@endlink[0]{}%
\providecommand \url  [0]{\begingroup\@sanitize@url \@url }%
\providecommand \@url [1]{\endgroup\@href {#1}{\urlprefix }}%
\providecommand \urlprefix  [0]{URL }%
\providecommand \Eprint [0]{\href }%
\providecommand \doibase [0]{http://dx.doi.org/}%
\providecommand \selectlanguage [0]{\@gobble}%
\providecommand \bibinfo  [0]{\@secondoftwo}%
\providecommand \bibfield  [0]{\@secondoftwo}%
\providecommand \translation [1]{[#1]}%
\providecommand \BibitemOpen [0]{}%
\providecommand \bibitemStop [0]{}%
\providecommand \bibitemNoStop [0]{.\EOS\space}%
\providecommand \EOS [0]{\spacefactor3000\relax}%
\providecommand \BibitemShut  [1]{\csname bibitem#1\endcsname}%
\let\auto@bib@innerbib\@empty
%</preamble>
\bibitem [{\citenamefont {Campbell}(1998)}]{campbell1998surface}%
  \BibitemOpen
  \bibfield  {author} {\bibinfo {author} {\bibfnamefont {C.}~\bibnamefont {Campbell}},\ }\href@noop {} {\emph {\bibinfo {title} {Surface Acoustic Wave Devices for Mobile and Wireless Communications, Four-Volume Set}}}\ (\bibinfo  {publisher} {Academic press},\ \bibinfo {year} {1998})\BibitemShut {NoStop}%
\bibitem [{\citenamefont {Ruby}(2015)}]{7153041}%
  \BibitemOpen
  \bibfield  {author} {\bibinfo {author} {\bibfnamefont {R.}~\bibnamefont {Ruby}},\ }\href {\doibase 10.1109/MMM.2015.2429513} {\bibfield  {journal} {\bibinfo  {journal} {IEEE Microwave Magazine}\ }\textbf {\bibinfo {volume} {16}},\ \bibinfo {pages} {46} (\bibinfo {year} {2015})}\BibitemShut {NoStop}%
\bibitem [{\citenamefont {Gronewold}(2007)}]{gronewold2007surface}%
  \BibitemOpen
  \bibfield  {author} {\bibinfo {author} {\bibfnamefont {T.~M.}\ \bibnamefont {Gronewold}},\ }\href {\doibase https://doi.org/10.1016/j.aca.2007.09.056} {\bibfield  {journal} {\bibinfo  {journal} {Analytica Chimica Acta}\ }\textbf {\bibinfo {volume} {603}},\ \bibinfo {pages} {119} (\bibinfo {year} {2007})}\BibitemShut {NoStop}%
\bibitem [{\citenamefont {Mandal}\ and\ \citenamefont {Banerjee}(2022)}]{mandal2022surface}%
  \BibitemOpen
  \bibfield  {author} {\bibinfo {author} {\bibfnamefont {D.}~\bibnamefont {Mandal}}\ and\ \bibinfo {author} {\bibfnamefont {S.}~\bibnamefont {Banerjee}},\ }\href {\doibase 10.3390/s22030820} {\bibfield  {journal} {\bibinfo  {journal} {Sensors}\ }\textbf {\bibinfo {volume} {22}} (\bibinfo {year} {2022}),\ 10.3390/s22030820}\BibitemShut {NoStop}%
\bibitem [{\citenamefont {Wohltjen}(1984)}]{wohltjen1984mechanism}%
  \BibitemOpen
  \bibfield  {author} {\bibinfo {author} {\bibfnamefont {H.}~\bibnamefont {Wohltjen}},\ }\href {\doibase https://doi.org/10.1016/0250-6874(84)85014-3} {\bibfield  {journal} {\bibinfo  {journal} {Sensors and Actuators}\ }\textbf {\bibinfo {volume} {5}},\ \bibinfo {pages} {307} (\bibinfo {year} {1984})}\BibitemShut {NoStop}%
\bibitem [{\citenamefont {Parker}\ and\ \citenamefont {Montress}(1988)}]{parker1988precision}%
  \BibitemOpen
  \bibfield  {author} {\bibinfo {author} {\bibfnamefont {T.}~\bibnamefont {Parker}}\ and\ \bibinfo {author} {\bibfnamefont {G.}~\bibnamefont {Montress}},\ }\href {\doibase 10.1109/58.20455} {\bibfield  {journal} {\bibinfo  {journal} {IEEE Transactions on Ultrasonics, Ferroelectrics, and Frequency Control}\ }\textbf {\bibinfo {volume} {35}},\ \bibinfo {pages} {342} (\bibinfo {year} {1988})}\BibitemShut {NoStop}%
\bibitem [{\citenamefont {Ding}\ \emph {et~al.}(2013)\citenamefont {Ding}, \citenamefont {Li}, \citenamefont {Lin}, \citenamefont {Stratton}, \citenamefont {Nama}, \citenamefont {Guo}, \citenamefont {Slotcavage}, \citenamefont {Mao}, \citenamefont {Shi}, \citenamefont {Costanzo} \emph {et~al.}}]{ding2013surface}%
  \BibitemOpen
  \bibfield  {author} {\bibinfo {author} {\bibfnamefont {X.}~\bibnamefont {Ding}}, \bibinfo {author} {\bibfnamefont {P.}~\bibnamefont {Li}}, \bibinfo {author} {\bibfnamefont {S.-C.~S.}\ \bibnamefont {Lin}}, \bibinfo {author} {\bibfnamefont {Z.~S.}\ \bibnamefont {Stratton}}, \bibinfo {author} {\bibfnamefont {N.}~\bibnamefont {Nama}}, \bibinfo {author} {\bibfnamefont {F.}~\bibnamefont {Guo}}, \bibinfo {author} {\bibfnamefont {D.}~\bibnamefont {Slotcavage}}, \bibinfo {author} {\bibfnamefont {X.}~\bibnamefont {Mao}}, \bibinfo {author} {\bibfnamefont {J.}~\bibnamefont {Shi}}, \bibinfo {author} {\bibfnamefont {F.}~\bibnamefont {Costanzo}},  \emph {et~al.},\ }\href {\doibase 10.1039/C3LC50361E} {\bibfield  {journal} {\bibinfo  {journal} {Lab on a Chip}\ }\textbf {\bibinfo {volume} {13}},\ \bibinfo {pages} {3626} (\bibinfo {year} {2013})}\BibitemShut {NoStop}%
\bibitem [{\citenamefont {Destgeer}\ and\ \citenamefont {Sung}(2015)}]{destgeer2015recent}%
  \BibitemOpen
  \bibfield  {author} {\bibinfo {author} {\bibfnamefont {G.}~\bibnamefont {Destgeer}}\ and\ \bibinfo {author} {\bibfnamefont {H.~J.}\ \bibnamefont {Sung}},\ }\href {\doibase 10.1039/C5LC00265F} {\bibfield  {journal} {\bibinfo  {journal} {Lab on a Chip}\ }\textbf {\bibinfo {volume} {15}},\ \bibinfo {pages} {2722} (\bibinfo {year} {2015})}\BibitemShut {NoStop}%
\bibitem [{\citenamefont {Sasaki}\ \emph {et~al.}(2017)\citenamefont {Sasaki}, \citenamefont {Nii}, \citenamefont {Iguchi},\ and\ \citenamefont {Onose}}]{sasaki2017nonreciprocal}%
  \BibitemOpen
  \bibfield  {author} {\bibinfo {author} {\bibfnamefont {R.}~\bibnamefont {Sasaki}}, \bibinfo {author} {\bibfnamefont {Y.}~\bibnamefont {Nii}}, \bibinfo {author} {\bibfnamefont {Y.}~\bibnamefont {Iguchi}}, \ and\ \bibinfo {author} {\bibfnamefont {Y.}~\bibnamefont {Onose}},\ }\href {\doibase 10.1103/PhysRevB.95.020407} {\bibfield  {journal} {\bibinfo  {journal} {Phys. Rev. B}\ }\textbf {\bibinfo {volume} {95}},\ \bibinfo {pages} {020407(R)} (\bibinfo {year} {2017})}\BibitemShut {NoStop}%
\bibitem [{\citenamefont {Tateno}\ and\ \citenamefont {Nozaki}(2020)}]{tateno2020highly}%
  \BibitemOpen
  \bibfield  {author} {\bibinfo {author} {\bibfnamefont {S.}~\bibnamefont {Tateno}}\ and\ \bibinfo {author} {\bibfnamefont {Y.}~\bibnamefont {Nozaki}},\ }\href {\doibase 10.1103/PhysRevApplied.13.034074} {\bibfield  {journal} {\bibinfo  {journal} {Phys. Rev. Appl.}\ }\textbf {\bibinfo {volume} {13}},\ \bibinfo {pages} {034074} (\bibinfo {year} {2020})}\BibitemShut {NoStop}%
\bibitem [{\citenamefont {Hern\'andez-M\'{\i}nguez}\ \emph {et~al.}(2020)\citenamefont {Hern\'andez-M\'{\i}nguez}, \citenamefont {Maci\`a}, \citenamefont {Hern\`andez}, \citenamefont {Herfort},\ and\ \citenamefont {Santos}}]{hernandez2020large}%
  \BibitemOpen
  \bibfield  {author} {\bibinfo {author} {\bibfnamefont {A.}~\bibnamefont {Hern\'andez-M\'{\i}nguez}}, \bibinfo {author} {\bibfnamefont {F.}~\bibnamefont {Maci\`a}}, \bibinfo {author} {\bibfnamefont {J.~M.}\ \bibnamefont {Hern\`andez}}, \bibinfo {author} {\bibfnamefont {J.}~\bibnamefont {Herfort}}, \ and\ \bibinfo {author} {\bibfnamefont {P.~V.}\ \bibnamefont {Santos}},\ }\href {\doibase 10.1103/PhysRevApplied.13.044018} {\bibfield  {journal} {\bibinfo  {journal} {Phys. Rev. Appl.}\ }\textbf {\bibinfo {volume} {13}},\ \bibinfo {pages} {044018} (\bibinfo {year} {2020})}\BibitemShut {NoStop}%
\bibitem [{\citenamefont {Xu}\ \emph {et~al.}(2020)\citenamefont {Xu}, \citenamefont {Yamamoto}, \citenamefont {Puebla}, \citenamefont {Baumgaertl}, \citenamefont {Rana}, \citenamefont {Miura}, \citenamefont {Takahashi}, \citenamefont {Grundler}, \citenamefont {Maekawa},\ and\ \citenamefont {Otani}}]{xu2020nonreciprocal}%
  \BibitemOpen
  \bibfield  {author} {\bibinfo {author} {\bibfnamefont {M.}~\bibnamefont {Xu}}, \bibinfo {author} {\bibfnamefont {K.}~\bibnamefont {Yamamoto}}, \bibinfo {author} {\bibfnamefont {J.}~\bibnamefont {Puebla}}, \bibinfo {author} {\bibfnamefont {K.}~\bibnamefont {Baumgaertl}}, \bibinfo {author} {\bibfnamefont {B.}~\bibnamefont {Rana}}, \bibinfo {author} {\bibfnamefont {K.}~\bibnamefont {Miura}}, \bibinfo {author} {\bibfnamefont {H.}~\bibnamefont {Takahashi}}, \bibinfo {author} {\bibfnamefont {D.}~\bibnamefont {Grundler}}, \bibinfo {author} {\bibfnamefont {S.}~\bibnamefont {Maekawa}}, \ and\ \bibinfo {author} {\bibfnamefont {Y.}~\bibnamefont {Otani}},\ }\href {\doibase 10.1126/sciadv.abb1724} {\bibfield  {journal} {\bibinfo  {journal} {Science Advances}\ }\textbf {\bibinfo {volume} {6}},\ \bibinfo {pages} {eabb1724} (\bibinfo {year} {2020})}\BibitemShut {NoStop}%
\bibitem [{\citenamefont {K\"u\ss{}}\ \emph {et~al.}(2020)\citenamefont {K\"u\ss{}}, \citenamefont {Heigl}, \citenamefont {Flacke}, \citenamefont {H\"orner}, \citenamefont {Weiler}, \citenamefont {Albrecht},\ and\ \citenamefont {Wixforth}}]{kuss2020nonreciprocal}%
  \BibitemOpen
  \bibfield  {author} {\bibinfo {author} {\bibfnamefont {M.}~\bibnamefont {K\"u\ss{}}}, \bibinfo {author} {\bibfnamefont {M.}~\bibnamefont {Heigl}}, \bibinfo {author} {\bibfnamefont {L.}~\bibnamefont {Flacke}}, \bibinfo {author} {\bibfnamefont {A.}~\bibnamefont {H\"orner}}, \bibinfo {author} {\bibfnamefont {M.}~\bibnamefont {Weiler}}, \bibinfo {author} {\bibfnamefont {M.}~\bibnamefont {Albrecht}}, \ and\ \bibinfo {author} {\bibfnamefont {A.}~\bibnamefont {Wixforth}},\ }\href {\doibase 10.1103/PhysRevLett.125.217203} {\bibfield  {journal} {\bibinfo  {journal} {Phys. Rev. Lett.}\ }\textbf {\bibinfo {volume} {125}},\ \bibinfo {pages} {217203} (\bibinfo {year} {2020})}\BibitemShut {NoStop}%
\bibitem [{\citenamefont {K\"u\ss{}}\ \emph {et~al.}(2021{\natexlab{a}})\citenamefont {K\"u\ss{}}, \citenamefont {Heigl}, \citenamefont {Flacke}, \citenamefont {Hefele}, \citenamefont {H\"orner}, \citenamefont {Weiler}, \citenamefont {Albrecht},\ and\ \citenamefont {Wixforth}}]{kuss2021symmetry}%
  \BibitemOpen
  \bibfield  {author} {\bibinfo {author} {\bibfnamefont {M.}~\bibnamefont {K\"u\ss{}}}, \bibinfo {author} {\bibfnamefont {M.}~\bibnamefont {Heigl}}, \bibinfo {author} {\bibfnamefont {L.}~\bibnamefont {Flacke}}, \bibinfo {author} {\bibfnamefont {A.}~\bibnamefont {Hefele}}, \bibinfo {author} {\bibfnamefont {A.}~\bibnamefont {H\"orner}}, \bibinfo {author} {\bibfnamefont {M.}~\bibnamefont {Weiler}}, \bibinfo {author} {\bibfnamefont {M.}~\bibnamefont {Albrecht}}, \ and\ \bibinfo {author} {\bibfnamefont {A.}~\bibnamefont {Wixforth}},\ }\href {\doibase 10.1103/PhysRevApplied.15.034046} {\bibfield  {journal} {\bibinfo  {journal} {Phys. Rev. Appl.}\ }\textbf {\bibinfo {volume} {15}},\ \bibinfo {pages} {034046} (\bibinfo {year} {2021}{\natexlab{a}})}\BibitemShut {NoStop}%
\bibitem [{\citenamefont {K\"u\ss{}}\ \emph {et~al.}(2021{\natexlab{b}})\citenamefont {K\"u\ss{}}, \citenamefont {Heigl}, \citenamefont {Flacke}, \citenamefont {H\"orner}, \citenamefont {Weiler}, \citenamefont {Wixforth},\ and\ \citenamefont {Albrecht}}]{kuss2021nonreciprocal}%
  \BibitemOpen
  \bibfield  {author} {\bibinfo {author} {\bibfnamefont {M.}~\bibnamefont {K\"u\ss{}}}, \bibinfo {author} {\bibfnamefont {M.}~\bibnamefont {Heigl}}, \bibinfo {author} {\bibfnamefont {L.}~\bibnamefont {Flacke}}, \bibinfo {author} {\bibfnamefont {A.}~\bibnamefont {H\"orner}}, \bibinfo {author} {\bibfnamefont {M.}~\bibnamefont {Weiler}}, \bibinfo {author} {\bibfnamefont {A.}~\bibnamefont {Wixforth}}, \ and\ \bibinfo {author} {\bibfnamefont {M.}~\bibnamefont {Albrecht}},\ }\href {\doibase 10.1103/PhysRevApplied.15.034060} {\bibfield  {journal} {\bibinfo  {journal} {Phys. Rev. Appl.}\ }\textbf {\bibinfo {volume} {15}},\ \bibinfo {pages} {034060} (\bibinfo {year} {2021}{\natexlab{b}})}\BibitemShut {NoStop}%
\bibitem [{\citenamefont {Li}\ \emph {et~al.}(2021)\citenamefont {Li}, \citenamefont {Zhao}, \citenamefont {Zhang}, \citenamefont {Hoffmann},\ and\ \citenamefont {Novosad}}]{li2021advances}%
  \BibitemOpen
  \bibfield  {author} {\bibinfo {author} {\bibfnamefont {Y.}~\bibnamefont {Li}}, \bibinfo {author} {\bibfnamefont {C.}~\bibnamefont {Zhao}}, \bibinfo {author} {\bibfnamefont {W.}~\bibnamefont {Zhang}}, \bibinfo {author} {\bibfnamefont {A.}~\bibnamefont {Hoffmann}}, \ and\ \bibinfo {author} {\bibfnamefont {V.}~\bibnamefont {Novosad}},\ }\href {\doibase 10.1063/5.0047054} {\bibfield  {journal} {\bibinfo  {journal} {APL Materials}\ }\textbf {\bibinfo {volume} {9}},\ \bibinfo {pages} {060902} (\bibinfo {year} {2021})}\BibitemShut {NoStop}%
\bibitem [{\citenamefont {Geilen}\ \emph {et~al.}(2022{\natexlab{a}})\citenamefont {Geilen}, \citenamefont {Nicoloiu}, \citenamefont {Narducci}, \citenamefont {Mohseni}, \citenamefont {Bechberger}, \citenamefont {Ender}, \citenamefont {Ciubotaru}, \citenamefont {Hillebrands}, \citenamefont {Müller}, \citenamefont {Adelmann},\ and\ \citenamefont {Pirro}}]{geilen2022fully}%
  \BibitemOpen
  \bibfield  {author} {\bibinfo {author} {\bibfnamefont {M.}~\bibnamefont {Geilen}}, \bibinfo {author} {\bibfnamefont {A.}~\bibnamefont {Nicoloiu}}, \bibinfo {author} {\bibfnamefont {D.}~\bibnamefont {Narducci}}, \bibinfo {author} {\bibfnamefont {M.}~\bibnamefont {Mohseni}}, \bibinfo {author} {\bibfnamefont {M.}~\bibnamefont {Bechberger}}, \bibinfo {author} {\bibfnamefont {M.}~\bibnamefont {Ender}}, \bibinfo {author} {\bibfnamefont {F.}~\bibnamefont {Ciubotaru}}, \bibinfo {author} {\bibfnamefont {B.}~\bibnamefont {Hillebrands}}, \bibinfo {author} {\bibfnamefont {A.}~\bibnamefont {Müller}}, \bibinfo {author} {\bibfnamefont {C.}~\bibnamefont {Adelmann}}, \ and\ \bibinfo {author} {\bibfnamefont {P.}~\bibnamefont {Pirro}},\ }\href {\doibase 10.1063/5.0088924} {\bibfield  {journal} {\bibinfo  {journal} {Applied Physics Letters}\ }\textbf {\bibinfo {volume} {120}},\ \bibinfo {pages} {242404} (\bibinfo {year} {2022}{\natexlab{a}})}\BibitemShut {NoStop}%
\bibitem [{\citenamefont {K\"u\ss{}}\ \emph {et~al.}(2023{\natexlab{a}})\citenamefont {K\"u\ss{}}, \citenamefont {Hassan}, \citenamefont {Kunz}, \citenamefont {H\"orner}, \citenamefont {Weiler},\ and\ \citenamefont {Albrecht}}]{kuss2023nonreciprocal1}%
  \BibitemOpen
  \bibfield  {author} {\bibinfo {author} {\bibfnamefont {M.}~\bibnamefont {K\"u\ss{}}}, \bibinfo {author} {\bibfnamefont {M.}~\bibnamefont {Hassan}}, \bibinfo {author} {\bibfnamefont {Y.}~\bibnamefont {Kunz}}, \bibinfo {author} {\bibfnamefont {A.}~\bibnamefont {H\"orner}}, \bibinfo {author} {\bibfnamefont {M.}~\bibnamefont {Weiler}}, \ and\ \bibinfo {author} {\bibfnamefont {M.}~\bibnamefont {Albrecht}},\ }\href {\doibase 10.1103/PhysRevB.107.024424} {\bibfield  {journal} {\bibinfo  {journal} {Phys. Rev. B}\ }\textbf {\bibinfo {volume} {107}},\ \bibinfo {pages} {024424} (\bibinfo {year} {2023}{\natexlab{a}})}\BibitemShut {NoStop}%
\bibitem [{\citenamefont {K\"u\ss{}}\ \emph {et~al.}(2023{\natexlab{b}})\citenamefont {K\"u\ss{}}, \citenamefont {Hassan}, \citenamefont {Kunz}, \citenamefont {H\"orner}, \citenamefont {Weiler},\ and\ \citenamefont {Albrecht}}]{kuss2023nonreciprocal}%
  \BibitemOpen
  \bibfield  {author} {\bibinfo {author} {\bibfnamefont {M.}~\bibnamefont {K\"u\ss{}}}, \bibinfo {author} {\bibfnamefont {M.}~\bibnamefont {Hassan}}, \bibinfo {author} {\bibfnamefont {Y.}~\bibnamefont {Kunz}}, \bibinfo {author} {\bibfnamefont {A.}~\bibnamefont {H\"orner}}, \bibinfo {author} {\bibfnamefont {M.}~\bibnamefont {Weiler}}, \ and\ \bibinfo {author} {\bibfnamefont {M.}~\bibnamefont {Albrecht}},\ }\href {\doibase 10.1103/PhysRevB.107.214412} {\bibfield  {journal} {\bibinfo  {journal} {Phys. Rev. B}\ }\textbf {\bibinfo {volume} {107}},\ \bibinfo {pages} {214412} (\bibinfo {year} {2023}{\natexlab{b}})}\BibitemShut {NoStop}%
\bibitem [{\citenamefont {Huang}\ \emph {et~al.}(2023)\citenamefont {Huang}, \citenamefont {Hu}, \citenamefont {Zhang},\ and\ \citenamefont {Bai}}]{huang2023phonon}%
  \BibitemOpen
  \bibfield  {author} {\bibinfo {author} {\bibfnamefont {M.}~\bibnamefont {Huang}}, \bibinfo {author} {\bibfnamefont {W.}~\bibnamefont {Hu}}, \bibinfo {author} {\bibfnamefont {H.}~\bibnamefont {Zhang}}, \ and\ \bibinfo {author} {\bibfnamefont {F.}~\bibnamefont {Bai}},\ }\href {\doibase 10.1063/5.0151667} {\bibfield  {journal} {\bibinfo  {journal} {Journal of Applied Physics}\ }\textbf {\bibinfo {volume} {133}},\ \bibinfo {pages} {223902} (\bibinfo {year} {2023})}\BibitemShut {NoStop}%
\bibitem [{\citenamefont {Kunz}\ \emph {et~al.}(2023)\citenamefont {Kunz}, \citenamefont {Küß}, \citenamefont {Schneider}, \citenamefont {Geilen}, \citenamefont {Pirro}, \citenamefont {Albrecht},\ and\ \citenamefont {Weiler}}]{kunz2023coherent}%
  \BibitemOpen
  \bibfield  {author} {\bibinfo {author} {\bibfnamefont {Y.}~\bibnamefont {Kunz}}, \bibinfo {author} {\bibfnamefont {M.}~\bibnamefont {Küß}}, \bibinfo {author} {\bibfnamefont {M.}~\bibnamefont {Schneider}}, \bibinfo {author} {\bibfnamefont {M.}~\bibnamefont {Geilen}}, \bibinfo {author} {\bibfnamefont {P.}~\bibnamefont {Pirro}}, \bibinfo {author} {\bibfnamefont {M.}~\bibnamefont {Albrecht}}, \ and\ \bibinfo {author} {\bibfnamefont {M.}~\bibnamefont {Weiler}},\ }\href@noop {} {\bibfield  {journal} {\bibinfo  {journal} {arXiv preprint arXiv:2311.16688}\ } (\bibinfo {year} {2023})},\ \Eprint {http://arxiv.org/abs/2311.16688} {arXiv:2311.16688} \BibitemShut {NoStop}%
\bibitem [{\citenamefont {Shah}\ \emph {et~al.}(2020)\citenamefont {Shah}, \citenamefont {Bas}, \citenamefont {Lisenkov}, \citenamefont {Matyushov}, \citenamefont {Sun},\ and\ \citenamefont {Page}}]{shah2020giant}%
  \BibitemOpen
  \bibfield  {author} {\bibinfo {author} {\bibfnamefont {P.~J.}\ \bibnamefont {Shah}}, \bibinfo {author} {\bibfnamefont {D.~A.}\ \bibnamefont {Bas}}, \bibinfo {author} {\bibfnamefont {I.}~\bibnamefont {Lisenkov}}, \bibinfo {author} {\bibfnamefont {A.}~\bibnamefont {Matyushov}}, \bibinfo {author} {\bibfnamefont {N.~X.}\ \bibnamefont {Sun}}, \ and\ \bibinfo {author} {\bibfnamefont {M.~R.}\ \bibnamefont {Page}},\ }\href {\doibase 10.1126/sciadv.abc5648} {\bibfield  {journal} {\bibinfo  {journal} {Science Advances}\ }\textbf {\bibinfo {volume} {6}},\ \bibinfo {pages} {eabc5648} (\bibinfo {year} {2020})}\BibitemShut {NoStop}%
\bibitem [{\citenamefont {Verba}\ \emph {et~al.}(2021)\citenamefont {Verba}, \citenamefont {Bankowski}, \citenamefont {Meitzler}, \citenamefont {Tiberkevich},\ and\ \citenamefont {Slavin}}]{verba2021phase}%
  \BibitemOpen
  \bibfield  {author} {\bibinfo {author} {\bibfnamefont {R.}~\bibnamefont {Verba}}, \bibinfo {author} {\bibfnamefont {E.~N.}\ \bibnamefont {Bankowski}}, \bibinfo {author} {\bibfnamefont {T.~J.}\ \bibnamefont {Meitzler}}, \bibinfo {author} {\bibfnamefont {V.}~\bibnamefont {Tiberkevich}}, \ and\ \bibinfo {author} {\bibfnamefont {A.}~\bibnamefont {Slavin}},\ }\href {\doibase 10.1002/aelm.202100263} {\bibfield  {journal} {\bibinfo  {journal} {Advanced Electronic Materials}\ }\textbf {\bibinfo {volume} {7}},\ \bibinfo {pages} {2100263} (\bibinfo {year} {2021})}\BibitemShut {NoStop}%
\bibitem [{\citenamefont {Rasmussen}\ \emph {et~al.}(2021)\citenamefont {Rasmussen}, \citenamefont {Quan},\ and\ \citenamefont {Alù}}]{rasmussen2021acoustic}%
  \BibitemOpen
  \bibfield  {author} {\bibinfo {author} {\bibfnamefont {C.}~\bibnamefont {Rasmussen}}, \bibinfo {author} {\bibfnamefont {L.}~\bibnamefont {Quan}}, \ and\ \bibinfo {author} {\bibfnamefont {A.}~\bibnamefont {Alù}},\ }\href {\doibase 10.1063/5.0050775} {\bibfield  {journal} {\bibinfo  {journal} {Journal of Applied Physics}\ }\textbf {\bibinfo {volume} {129}},\ \bibinfo {pages} {210903} (\bibinfo {year} {2021})}\BibitemShut {NoStop}%
\bibitem [{\citenamefont {Küß}\ \emph {et~al.}(2022)\citenamefont {Küß}, \citenamefont {Albrecht},\ and\ \citenamefont {Weiler}}]{kuss2022chiral}%
  \BibitemOpen
  \bibfield  {author} {\bibinfo {author} {\bibfnamefont {M.}~\bibnamefont {Küß}}, \bibinfo {author} {\bibfnamefont {M.}~\bibnamefont {Albrecht}}, \ and\ \bibinfo {author} {\bibfnamefont {M.}~\bibnamefont {Weiler}},\ }\href {\doibase 10.3389/fphy.2022.981257} {\bibfield  {journal} {\bibinfo  {journal} {Frontiers in Physics}\ }\textbf {\bibinfo {volume} {10}} (\bibinfo {year} {2022}),\ 10.3389/fphy.2022.981257}\BibitemShut {NoStop}%
\bibitem [{\citenamefont {Li}\ \emph {et~al.}(2017)\citenamefont {Li}, \citenamefont {Labanowski}, \citenamefont {Salahuddin},\ and\ \citenamefont {Lynch}}]{li2017spin}%
  \BibitemOpen
  \bibfield  {author} {\bibinfo {author} {\bibfnamefont {X.}~\bibnamefont {Li}}, \bibinfo {author} {\bibfnamefont {D.}~\bibnamefont {Labanowski}}, \bibinfo {author} {\bibfnamefont {S.}~\bibnamefont {Salahuddin}}, \ and\ \bibinfo {author} {\bibfnamefont {C.~S.}\ \bibnamefont {Lynch}},\ }\href {\doibase 10.1063/1.4996102} {\bibfield  {journal} {\bibinfo  {journal} {Journal of Applied Physics}\ }\textbf {\bibinfo {volume} {122}},\ \bibinfo {pages} {043904} (\bibinfo {year} {2017})}\BibitemShut {NoStop}%
\bibitem [{\citenamefont {Mahmoud}\ \emph {et~al.}(2020)\citenamefont {Mahmoud}, \citenamefont {Ciubotaru}, \citenamefont {Vanderveken}, \citenamefont {Chumak}, \citenamefont {Hamdioui}, \citenamefont {Adelmann},\ and\ \citenamefont {Cotofana}}]{mahmoud2020introduction}%
  \BibitemOpen
  \bibfield  {author} {\bibinfo {author} {\bibfnamefont {A.}~\bibnamefont {Mahmoud}}, \bibinfo {author} {\bibfnamefont {F.}~\bibnamefont {Ciubotaru}}, \bibinfo {author} {\bibfnamefont {F.}~\bibnamefont {Vanderveken}}, \bibinfo {author} {\bibfnamefont {A.~V.}\ \bibnamefont {Chumak}}, \bibinfo {author} {\bibfnamefont {S.}~\bibnamefont {Hamdioui}}, \bibinfo {author} {\bibfnamefont {C.}~\bibnamefont {Adelmann}}, \ and\ \bibinfo {author} {\bibfnamefont {S.}~\bibnamefont {Cotofana}},\ }\href {\doibase 10.1063/5.0019328} {\bibfield  {journal} {\bibinfo  {journal} {Journal of Applied Physics}\ }\textbf {\bibinfo {volume} {128}},\ \bibinfo {pages} {161101} (\bibinfo {year} {2020})}\BibitemShut {NoStop}%
\bibitem [{\citenamefont {Chumak}\ \emph {et~al.}(2022)\citenamefont {Chumak}, \citenamefont {Kabos}, \citenamefont {Wu}, \citenamefont {Abert}, \citenamefont {Adelmann}, \citenamefont {Adeyeye}, \citenamefont {{\AA}kerman}, \citenamefont {Aliev}, \citenamefont {Anane}, \citenamefont {Awad} \emph {et~al.}}]{chumak2022advances}%
  \BibitemOpen
  \bibfield  {author} {\bibinfo {author} {\bibfnamefont {A.~V.}\ \bibnamefont {Chumak}}, \bibinfo {author} {\bibfnamefont {P.}~\bibnamefont {Kabos}}, \bibinfo {author} {\bibfnamefont {M.}~\bibnamefont {Wu}}, \bibinfo {author} {\bibfnamefont {C.}~\bibnamefont {Abert}}, \bibinfo {author} {\bibfnamefont {C.}~\bibnamefont {Adelmann}}, \bibinfo {author} {\bibfnamefont {A.}~\bibnamefont {Adeyeye}}, \bibinfo {author} {\bibfnamefont {J.}~\bibnamefont {{\AA}kerman}}, \bibinfo {author} {\bibfnamefont {F.~G.}\ \bibnamefont {Aliev}}, \bibinfo {author} {\bibfnamefont {A.}~\bibnamefont {Anane}}, \bibinfo {author} {\bibfnamefont {A.}~\bibnamefont {Awad}},  \emph {et~al.},\ }\href {\doibase 10.1109/TMAG.2022.3149664} {\bibfield  {journal} {\bibinfo  {journal} {IEEE Transactions on Magnetics}\ }\textbf {\bibinfo {volume} {58}},\ \bibinfo {pages} {1} (\bibinfo {year} {2022})}\BibitemShut {NoStop}%
\bibitem [{\citenamefont {Rana}\ \emph {et~al.}(2017)\citenamefont {Rana}, \citenamefont {Fukuma}, \citenamefont {Miura}, \citenamefont {Takahashi},\ and\ \citenamefont {Otani}}]{rana2017excitation}%
  \BibitemOpen
  \bibfield  {author} {\bibinfo {author} {\bibfnamefont {B.}~\bibnamefont {Rana}}, \bibinfo {author} {\bibfnamefont {Y.}~\bibnamefont {Fukuma}}, \bibinfo {author} {\bibfnamefont {K.}~\bibnamefont {Miura}}, \bibinfo {author} {\bibfnamefont {H.}~\bibnamefont {Takahashi}}, \ and\ \bibinfo {author} {\bibfnamefont {Y.}~\bibnamefont {Otani}},\ }\href {\doibase 10.1063/1.4990724} {\bibfield  {journal} {\bibinfo  {journal} {Applied Physics Letters}\ }\textbf {\bibinfo {volume} {111}},\ \bibinfo {pages} {052404} (\bibinfo {year} {2017})}\BibitemShut {NoStop}%
\bibitem [{\citenamefont {Nikitchenko}\ and\ \citenamefont {Pertsev}(2021)}]{nikitchenko2021spin}%
  \BibitemOpen
  \bibfield  {author} {\bibinfo {author} {\bibfnamefont {A.~I.}\ \bibnamefont {Nikitchenko}}\ and\ \bibinfo {author} {\bibfnamefont {N.~A.}\ \bibnamefont {Pertsev}},\ }\href {\doibase 10.1103/PhysRevB.104.134422} {\bibfield  {journal} {\bibinfo  {journal} {Phys. Rev. B}\ }\textbf {\bibinfo {volume} {104}},\ \bibinfo {pages} {134422} (\bibinfo {year} {2021})}\BibitemShut {NoStop}%
\bibitem [{\citenamefont {Qin}\ \emph {et~al.}(2018)\citenamefont {Qin}, \citenamefont {H\"am\"al\"ainen}, \citenamefont {Arjas}, \citenamefont {Witteveen},\ and\ \citenamefont {van Dijken}}]{qin2018propagating}%
  \BibitemOpen
  \bibfield  {author} {\bibinfo {author} {\bibfnamefont {H.}~\bibnamefont {Qin}}, \bibinfo {author} {\bibfnamefont {S.~J.}\ \bibnamefont {H\"am\"al\"ainen}}, \bibinfo {author} {\bibfnamefont {K.}~\bibnamefont {Arjas}}, \bibinfo {author} {\bibfnamefont {J.}~\bibnamefont {Witteveen}}, \ and\ \bibinfo {author} {\bibfnamefont {S.}~\bibnamefont {van Dijken}},\ }\href {\doibase 10.1103/PhysRevB.98.224422} {\bibfield  {journal} {\bibinfo  {journal} {Phys. Rev. B}\ }\textbf {\bibinfo {volume} {98}},\ \bibinfo {pages} {224422} (\bibinfo {year} {2018})}\BibitemShut {NoStop}%
\bibitem [{\citenamefont {Maendl}\ \emph {et~al.}(2017)\citenamefont {Maendl}, \citenamefont {Stasinopoulos},\ and\ \citenamefont {Grundler}}]{maendl2017spin}%
  \BibitemOpen
  \bibfield  {author} {\bibinfo {author} {\bibfnamefont {S.}~\bibnamefont {Maendl}}, \bibinfo {author} {\bibfnamefont {I.}~\bibnamefont {Stasinopoulos}}, \ and\ \bibinfo {author} {\bibfnamefont {D.}~\bibnamefont {Grundler}},\ }\href {\doibase 10.1063/1.4991520} {\bibfield  {journal} {\bibinfo  {journal} {Applied Physics Letters}\ }\textbf {\bibinfo {volume} {111}},\ \bibinfo {pages} {012403} (\bibinfo {year} {2017})}\BibitemShut {NoStop}%
\bibitem [{\citenamefont {Koike}\ \emph {et~al.}(1993)\citenamefont {Koike}, \citenamefont {Shimoe},\ and\ \citenamefont {Ieki}}]{koike19931}%
  \BibitemOpen
  \bibfield  {author} {\bibinfo {author} {\bibfnamefont {J.}~\bibnamefont {Koike}}, \bibinfo {author} {\bibfnamefont {K.~S.~K.}\ \bibnamefont {Shimoe}}, \ and\ \bibinfo {author} {\bibfnamefont {H.~I.~H.}\ \bibnamefont {Ieki}},\ }\href {\doibase 10.1143/JJAP.32.2337} {\bibfield  {journal} {\bibinfo  {journal} {Japanese Journal of Applied Physics}\ }\textbf {\bibinfo {volume} {32}},\ \bibinfo {pages} {2337} (\bibinfo {year} {1993})}\BibitemShut {NoStop}%
\bibitem [{\citenamefont {Le~Brizoual}\ \emph {et~al.}(2008)\citenamefont {Le~Brizoual}, \citenamefont {Sarry}, \citenamefont {Elmazria}, \citenamefont {Alnot}, \citenamefont {Ballandras},\ and\ \citenamefont {Pastureaud}}]{le2008ghz}%
  \BibitemOpen
  \bibfield  {author} {\bibinfo {author} {\bibfnamefont {L.}~\bibnamefont {Le~Brizoual}}, \bibinfo {author} {\bibfnamefont {F.}~\bibnamefont {Sarry}}, \bibinfo {author} {\bibfnamefont {O.}~\bibnamefont {Elmazria}}, \bibinfo {author} {\bibfnamefont {P.}~\bibnamefont {Alnot}}, \bibinfo {author} {\bibfnamefont {S.}~\bibnamefont {Ballandras}}, \ and\ \bibinfo {author} {\bibfnamefont {T.}~\bibnamefont {Pastureaud}},\ }\href {\doibase 10.1109/TUFFC.2008.662} {\bibfield  {journal} {\bibinfo  {journal} {IEEE Transactions on Ultrasonics, Ferroelectrics, and Frequency Control}\ }\textbf {\bibinfo {volume} {55}},\ \bibinfo {pages} {442} (\bibinfo {year} {2008})}\BibitemShut {NoStop}%
\bibitem [{\citenamefont {Wang}\ \emph {et~al.}(2008)\citenamefont {Wang}, \citenamefont {Pflügl}, \citenamefont {Andress}, \citenamefont {Ham}, \citenamefont {Capasso},\ and\ \citenamefont {Yamanishi}}]{wang2008gigahertz}%
  \BibitemOpen
  \bibfield  {author} {\bibinfo {author} {\bibfnamefont {Q.~J.}\ \bibnamefont {Wang}}, \bibinfo {author} {\bibfnamefont {C.}~\bibnamefont {Pflügl}}, \bibinfo {author} {\bibfnamefont {W.~F.}\ \bibnamefont {Andress}}, \bibinfo {author} {\bibfnamefont {D.}~\bibnamefont {Ham}}, \bibinfo {author} {\bibfnamefont {F.}~\bibnamefont {Capasso}}, \ and\ \bibinfo {author} {\bibfnamefont {M.}~\bibnamefont {Yamanishi}},\ }\href {\doibase 10.1116/1.2993176} {\bibfield  {journal} {\bibinfo  {journal} {Journal of Vacuum Science \& Technology B: Microelectronics and Nanometer Structures Processing, Measurement, and Phenomena}\ }\textbf {\bibinfo {volume} {26}},\ \bibinfo {pages} {1848} (\bibinfo {year} {2008})}\BibitemShut {NoStop}%
\bibitem [{\citenamefont {Fu}\ \emph {et~al.}(2019)\citenamefont {Fu}, \citenamefont {Wang}, \citenamefont {Qian}, \citenamefont {Li}, \citenamefont {Lu}, \citenamefont {Shen}, \citenamefont {Song}, \citenamefont {Zeng},\ and\ \citenamefont {Pan}}]{fu2018high}%
  \BibitemOpen
  \bibfield  {author} {\bibinfo {author} {\bibfnamefont {S.}~\bibnamefont {Fu}}, \bibinfo {author} {\bibfnamefont {W.}~\bibnamefont {Wang}}, \bibinfo {author} {\bibfnamefont {L.}~\bibnamefont {Qian}}, \bibinfo {author} {\bibfnamefont {Q.}~\bibnamefont {Li}}, \bibinfo {author} {\bibfnamefont {Z.}~\bibnamefont {Lu}}, \bibinfo {author} {\bibfnamefont {J.}~\bibnamefont {Shen}}, \bibinfo {author} {\bibfnamefont {C.}~\bibnamefont {Song}}, \bibinfo {author} {\bibfnamefont {F.}~\bibnamefont {Zeng}}, \ and\ \bibinfo {author} {\bibfnamefont {F.}~\bibnamefont {Pan}},\ }\href {\doibase 10.1109/LED.2018.2881467} {\bibfield  {journal} {\bibinfo  {journal} {IEEE Electron Device Letters}\ }\textbf {\bibinfo {volume} {40}},\ \bibinfo {pages} {103} (\bibinfo {year} {2019})}\BibitemShut {NoStop}%
\bibitem [{\citenamefont {Su}\ \emph {et~al.}(2020)\citenamefont {Su}, \citenamefont {Fu}, \citenamefont {Shen}, \citenamefont {Chen}, \citenamefont {Lu}, \citenamefont {Yang}, \citenamefont {Wang}, \citenamefont {Zeng}, \citenamefont {Wang}, \citenamefont {Song} \emph {et~al.}}]{su2020enhanced}%
  \BibitemOpen
  \bibfield  {author} {\bibinfo {author} {\bibfnamefont {R.}~\bibnamefont {Su}}, \bibinfo {author} {\bibfnamefont {S.}~\bibnamefont {Fu}}, \bibinfo {author} {\bibfnamefont {J.}~\bibnamefont {Shen}}, \bibinfo {author} {\bibfnamefont {Z.}~\bibnamefont {Chen}}, \bibinfo {author} {\bibfnamefont {Z.}~\bibnamefont {Lu}}, \bibinfo {author} {\bibfnamefont {M.}~\bibnamefont {Yang}}, \bibinfo {author} {\bibfnamefont {R.}~\bibnamefont {Wang}}, \bibinfo {author} {\bibfnamefont {F.}~\bibnamefont {Zeng}}, \bibinfo {author} {\bibfnamefont {W.}~\bibnamefont {Wang}}, \bibinfo {author} {\bibfnamefont {C.}~\bibnamefont {Song}},  \emph {et~al.},\ }\href {\doibase 10.1021/acsami.0c12055} {\bibfield  {journal} {\bibinfo  {journal} {ACS Applied Materials \& Interfaces}\ }\textbf {\bibinfo {volume} {12}},\ \bibinfo {pages} {42378} (\bibinfo {year} {2020})}\BibitemShut {NoStop}%
\bibitem [{\citenamefont {Hanna}\ and\ \citenamefont {Murphy}(1988)}]{hanna1988interactions}%
  \BibitemOpen
  \bibfield  {author} {\bibinfo {author} {\bibfnamefont {S.}~\bibnamefont {Hanna}}\ and\ \bibinfo {author} {\bibfnamefont {G.}~\bibnamefont {Murphy}},\ }\href {\doibase 10.1109/20.92254} {\bibfield  {journal} {\bibinfo  {journal} {IEEE Transactions on Magnetics}\ }\textbf {\bibinfo {volume} {24}},\ \bibinfo {pages} {2814} (\bibinfo {year} {1988})}\BibitemShut {NoStop}%
\bibitem [{\citenamefont {Kryshtal}\ and\ \citenamefont {Medved}(2017)}]{kryshtal2017nonlinear}%
  \BibitemOpen
  \bibfield  {author} {\bibinfo {author} {\bibfnamefont {R.}~\bibnamefont {Kryshtal}}\ and\ \bibinfo {author} {\bibfnamefont {A.}~\bibnamefont {Medved}},\ }\href {\doibase 10.1088/1361-6463/aa93ba} {\bibfield  {journal} {\bibinfo  {journal} {Journal of Physics D: Applied Physics}\ }\textbf {\bibinfo {volume} {50}},\ \bibinfo {pages} {495004} (\bibinfo {year} {2017})}\BibitemShut {NoStop}%
\bibitem [{\citenamefont {FARNELL}\ and\ \citenamefont {ADLER}(1972)}]{FARNELL197235}%
  \BibitemOpen
  \bibfield  {author} {\bibinfo {author} {\bibfnamefont {G.}~\bibnamefont {FARNELL}}\ and\ \bibinfo {author} {\bibfnamefont {E.}~\bibnamefont {ADLER}}\ }(\bibinfo  {publisher} {Academic Press},\ \bibinfo {year} {1972})\ pp.\ \bibinfo {pages} {35--127}\BibitemShut {NoStop}%
\bibitem [{\citenamefont {Zhang}(2022)}]{zhang_2022_bulk}%
  \BibitemOpen
  \bibfield  {author} {\bibinfo {author} {\bibfnamefont {G.}~\bibnamefont {Zhang}},\ }\href@noop {} {\emph {\bibinfo {title} {Bulk and Surface Acoustic Waves}}}\ (\bibinfo  {publisher} {CRC Press},\ \bibinfo {year} {2022})\BibitemShut {NoStop}%
\bibitem [{\citenamefont {Dubs}\ \emph {et~al.}(2020)\citenamefont {Dubs}, \citenamefont {Surzhenko}, \citenamefont {Thomas}, \citenamefont {Osten}, \citenamefont {Schneider}, \citenamefont {Lenz}, \citenamefont {Grenzer}, \citenamefont {H\"ubner},\ and\ \citenamefont {Wendler}}]{dubs2020low}%
  \BibitemOpen
  \bibfield  {author} {\bibinfo {author} {\bibfnamefont {C.}~\bibnamefont {Dubs}}, \bibinfo {author} {\bibfnamefont {O.}~\bibnamefont {Surzhenko}}, \bibinfo {author} {\bibfnamefont {R.}~\bibnamefont {Thomas}}, \bibinfo {author} {\bibfnamefont {J.}~\bibnamefont {Osten}}, \bibinfo {author} {\bibfnamefont {T.}~\bibnamefont {Schneider}}, \bibinfo {author} {\bibfnamefont {K.}~\bibnamefont {Lenz}}, \bibinfo {author} {\bibfnamefont {J.}~\bibnamefont {Grenzer}}, \bibinfo {author} {\bibfnamefont {R.}~\bibnamefont {H\"ubner}}, \ and\ \bibinfo {author} {\bibfnamefont {E.}~\bibnamefont {Wendler}},\ }\href {\doibase 10.1103/PhysRevMaterials.4.024416} {\bibfield  {journal} {\bibinfo  {journal} {Phys. Rev. Mater.}\ }\textbf {\bibinfo {volume} {4}},\ \bibinfo {pages} {024416} (\bibinfo {year} {2020})}\BibitemShut {NoStop}%
\bibitem [{Sup()}]{SupplemetalMaterial}%
  \BibitemOpen
  \href@noop {} {}\bibinfo {note} {See Supplemental Material for information about sample preparation; material parameters used in calculations; calculations of the electromechanical coupling coefficients, displacements, and strains; rotation of crystallographic axes including the piezoelectric ZnO c-axis; and the modelling of the `dead layer', which includes Refs.~\cite{bachari1999structural, lin2004effect, howe2015pseudomorphic, zhang_2022_bulk, clark1961elastic, hasan2017yttrium, graham1970elastic, connelly2021complex, FARNELL197235, kuss2020nonreciprocal, fung2015phonon, den1999relation, foster1968cadmium, Yanagitani2011cAxisZZ,laude2003slowness, o2020slowness, maupin2007introduction}}\BibitemShut {NoStop}%
\bibitem [{\citenamefont {Büttner}\ \emph {et~al.}(2000)\citenamefont {Büttner}, \citenamefont {Bauer}, \citenamefont {Demokritov}, \citenamefont {Hillebrands}, \citenamefont {Kivshar}, \citenamefont {Grimalsky}, \citenamefont {Rapoport}, \citenamefont {Kostylev}, \citenamefont {Kalinikos},\ and\ \citenamefont {Slavin}}]{TimetaggerPaper}%
  \BibitemOpen
  \bibfield  {author} {\bibinfo {author} {\bibfnamefont {O.}~\bibnamefont {Büttner}}, \bibinfo {author} {\bibfnamefont {M.}~\bibnamefont {Bauer}}, \bibinfo {author} {\bibfnamefont {S.~O.}\ \bibnamefont {Demokritov}}, \bibinfo {author} {\bibfnamefont {B.}~\bibnamefont {Hillebrands}}, \bibinfo {author} {\bibfnamefont {Y.~S.}\ \bibnamefont {Kivshar}}, \bibinfo {author} {\bibfnamefont {V.}~\bibnamefont {Grimalsky}}, \bibinfo {author} {\bibfnamefont {Y.}~\bibnamefont {Rapoport}}, \bibinfo {author} {\bibfnamefont {M.~P.}\ \bibnamefont {Kostylev}}, \bibinfo {author} {\bibfnamefont {B.~A.}\ \bibnamefont {Kalinikos}}, \ and\ \bibinfo {author} {\bibfnamefont {A.~N.}\ \bibnamefont {Slavin}},\ }\href {\doibase 10.1063/1.373257} {\bibfield  {journal} {\bibinfo  {journal} {Journal of Applied Physics}\ }\textbf {\bibinfo {volume} {87}},\ \bibinfo {pages} {5088} (\bibinfo {year} {2000})}\BibitemShut {NoStop}%
\bibitem [{\citenamefont {Demidov}\ \emph {et~al.}(2004)\citenamefont {Demidov}, \citenamefont {Demokritov}, \citenamefont {Hillebrands}, \citenamefont {Laufenberg},\ and\ \citenamefont {Freitas}}]{demidov2004radiation}%
  \BibitemOpen
  \bibfield  {author} {\bibinfo {author} {\bibfnamefont {V.~E.}\ \bibnamefont {Demidov}}, \bibinfo {author} {\bibfnamefont {S.~O.}\ \bibnamefont {Demokritov}}, \bibinfo {author} {\bibfnamefont {B.}~\bibnamefont {Hillebrands}}, \bibinfo {author} {\bibfnamefont {M.}~\bibnamefont {Laufenberg}}, \ and\ \bibinfo {author} {\bibfnamefont {P.~P.}\ \bibnamefont {Freitas}},\ }\href {\doibase 10.1063/1.1803621} {\bibfield  {journal} {\bibinfo  {journal} {Applied Physics Letters}\ }\textbf {\bibinfo {volume} {85}},\ \bibinfo {pages} {2866} (\bibinfo {year} {2004})}\BibitemShut {NoStop}%
\bibitem [{\citenamefont {Sebastian}\ \emph {et~al.}(2015)\citenamefont {Sebastian}, \citenamefont {Schultheiss}, \citenamefont {Obry}, \citenamefont {Hillebrands},\ and\ \citenamefont {Schultheiss}}]{TimetaggerFigure}%
  \BibitemOpen
  \bibfield  {author} {\bibinfo {author} {\bibfnamefont {T.}~\bibnamefont {Sebastian}}, \bibinfo {author} {\bibfnamefont {K.}~\bibnamefont {Schultheiss}}, \bibinfo {author} {\bibfnamefont {B.}~\bibnamefont {Obry}}, \bibinfo {author} {\bibfnamefont {B.}~\bibnamefont {Hillebrands}}, \ and\ \bibinfo {author} {\bibfnamefont {H.}~\bibnamefont {Schultheiss}},\ }\href {https://www.frontiersin.org/articles/10.3389/fphy.2015.00035} {\bibfield  {journal} {\bibinfo  {journal} {Frontiers in Physics}\ }\textbf {\bibinfo {volume} {3}} (\bibinfo {year} {2015})}\BibitemShut {NoStop}%
\bibitem [{\citenamefont {Heinz}(2021)}]{heinz2021nano}%
  \BibitemOpen
  \bibfield  {author} {\bibinfo {author} {\bibfnamefont {B.~M.}\ \bibnamefont {Heinz}},\ }\emph {\bibinfo {title} {Nano-scaled yttrium iron garnet conduits for magnonic networks}},\ \href@noop {} {Ph.D. thesis},\ \bibinfo  {school} {Rheinland-Pfälzische Technische Universität Kaiserslautern-Landau} (\bibinfo {year} {2021})\BibitemShut {NoStop}%
\bibitem [{\citenamefont {Rayleigh}(1885)}]{rayleigh1885waves}%
  \BibitemOpen
  \bibfield  {author} {\bibinfo {author} {\bibfnamefont {L.}~\bibnamefont {Rayleigh}},\ }\href {\doibase 10.1112/plms/s1-17.1.4} {\bibfield  {journal} {\bibinfo  {journal} {Proceedings of the London mathematical Society}\ }\textbf {\bibinfo {volume} {1}},\ \bibinfo {pages} {4} (\bibinfo {year} {1885})}\BibitemShut {NoStop}%
\bibitem [{\citenamefont {Sezawa}(1927)}]{Sezawa1927DispersionOE}%
  \BibitemOpen
  \bibfield  {author} {\bibinfo {author} {\bibfnamefont {K.}~\bibnamefont {Sezawa}},\ }\href@noop {} {\bibfield  {journal} {\bibinfo  {journal} {Bull. Earthq. Res. Inst. Tokyo}\ }\textbf {\bibinfo {volume} {3}},\ \bibinfo {pages} {1} (\bibinfo {year} {1927})}\BibitemShut {NoStop}%
\bibitem [{\citenamefont {Love}(1911)}]{love1911some}%
  \BibitemOpen
  \bibfield  {author} {\bibinfo {author} {\bibfnamefont {A.~E.~H.}\ \bibnamefont {Love}},\ }\href@noop {} {\emph {\bibinfo {title} {Some Problems of Geodynamics: Being an Essay to which the Adams Prize in the University of Cambridge was Adjudged in 1911}}}\ (\bibinfo  {publisher} {University Press},\ \bibinfo {year} {1911})\BibitemShut {NoStop}%
\bibitem [{\citenamefont {Clark}\ and\ \citenamefont {Strakna}(2004)}]{clark1961elastic}%
  \BibitemOpen
  \bibfield  {author} {\bibinfo {author} {\bibfnamefont {A.~E.}\ \bibnamefont {Clark}}\ and\ \bibinfo {author} {\bibfnamefont {R.~E.}\ \bibnamefont {Strakna}},\ }\href {\doibase 10.1063/1.1736184} {\bibfield  {journal} {\bibinfo  {journal} {Journal of Applied Physics}\ }\textbf {\bibinfo {volume} {32}},\ \bibinfo {pages} {1172} (\bibinfo {year} {2004})}\BibitemShut {NoStop}%
\bibitem [{\citenamefont {Hasan}\ \emph {et~al.}(2017)\citenamefont {Hasan}, \citenamefont {Hamidon}, \citenamefont {Ismail}, \citenamefont {Osman},\ and\ \citenamefont {Ismail}}]{hasan2017yttrium}%
  \BibitemOpen
  \bibfield  {author} {\bibinfo {author} {\bibfnamefont {I.~H.}\ \bibnamefont {Hasan}}, \bibinfo {author} {\bibfnamefont {M.~N.}\ \bibnamefont {Hamidon}}, \bibinfo {author} {\bibfnamefont {I.}~\bibnamefont {Ismail}}, \bibinfo {author} {\bibfnamefont {R.}~\bibnamefont {Osman}}, \ and\ \bibinfo {author} {\bibfnamefont {A.}~\bibnamefont {Ismail}},\ }in\ \href {\doibase 10.1109/RSM.2017.8069168} {\emph {\bibinfo {booktitle} {2017 IEEE Regional Symposium on Micro and Nanoelectronics (RSM)}}}\ (\bibinfo {year} {2017})\ pp.\ \bibinfo {pages} {131--134}\BibitemShut {NoStop}%
\bibitem [{\citenamefont {Graham}\ and\ \citenamefont {Chang}(2003)}]{graham1970elastic}%
  \BibitemOpen
  \bibfield  {author} {\bibinfo {author} {\bibfnamefont {L.~J.}\ \bibnamefont {Graham}}\ and\ \bibinfo {author} {\bibfnamefont {R.}~\bibnamefont {Chang}},\ }\href {\doibase 10.1063/1.1659197} {\bibfield  {journal} {\bibinfo  {journal} {Journal of Applied Physics}\ }\textbf {\bibinfo {volume} {41}},\ \bibinfo {pages} {2247} (\bibinfo {year} {2003})}\BibitemShut {NoStop}%
\bibitem [{\citenamefont {Connelly}\ \emph {et~al.}(2021)\citenamefont {Connelly}, \citenamefont {Aquino}, \citenamefont {Robbins}, \citenamefont {Bernstein}, \citenamefont {Orlov}, \citenamefont {Porod},\ and\ \citenamefont {Chisum}}]{connelly2021complex}%
  \BibitemOpen
  \bibfield  {author} {\bibinfo {author} {\bibfnamefont {D.~A.}\ \bibnamefont {Connelly}}, \bibinfo {author} {\bibfnamefont {H.~R.~O.}\ \bibnamefont {Aquino}}, \bibinfo {author} {\bibfnamefont {M.}~\bibnamefont {Robbins}}, \bibinfo {author} {\bibfnamefont {G.~H.}\ \bibnamefont {Bernstein}}, \bibinfo {author} {\bibfnamefont {A.}~\bibnamefont {Orlov}}, \bibinfo {author} {\bibfnamefont {W.}~\bibnamefont {Porod}}, \ and\ \bibinfo {author} {\bibfnamefont {J.}~\bibnamefont {Chisum}},\ }\href {\doibase 10.1109/LMAG.2021.3132850} {\bibfield  {journal} {\bibinfo  {journal} {IEEE Magnetics Letters}\ }\textbf {\bibinfo {volume} {12}},\ \bibinfo {pages} {1} (\bibinfo {year} {2021})}\BibitemShut {NoStop}%
\bibitem [{\citenamefont {Fung}(2015)}]{fung2015phonon}%
  \BibitemOpen
  \bibfield  {author} {\bibinfo {author} {\bibfnamefont {T.}~\bibnamefont {Fung}},\ }\emph {\bibinfo {title} {Phonon magnonics}},\ \href@noop {} {Ph.D. thesis},\ \bibinfo  {school} {University of Oxford} (\bibinfo {year} {2015})\BibitemShut {NoStop}%
\bibitem [{\citenamefont {Den~Toonder}\ \emph {et~al.}(1999)\citenamefont {Den~Toonder}, \citenamefont {Van~Dommelen},\ and\ \citenamefont {Baaijens}}]{den1999relation}%
  \BibitemOpen
  \bibfield  {author} {\bibinfo {author} {\bibfnamefont {J.}~\bibnamefont {Den~Toonder}}, \bibinfo {author} {\bibfnamefont {J.}~\bibnamefont {Van~Dommelen}}, \ and\ \bibinfo {author} {\bibfnamefont {F.}~\bibnamefont {Baaijens}},\ }\href {\doibase 10.1088/0965-0393/7/6/301} {\bibfield  {journal} {\bibinfo  {journal} {Modelling and Simulation in Materials Science and Engineering}\ }\textbf {\bibinfo {volume} {7}},\ \bibinfo {pages} {909} (\bibinfo {year} {1999})}\BibitemShut {NoStop}%
\bibitem [{\citenamefont {Hwang}\ \emph {et~al.}(2023)\citenamefont {Hwang}, \citenamefont {Puebla}, \citenamefont {Kondou}, \citenamefont {Gonzalez-Ballestero}, \citenamefont {Isshiki}, \citenamefont {Munoz}, \citenamefont {Liao}, \citenamefont {Chen}, \citenamefont {Luo}, \citenamefont {Maekawa} \emph {et~al.}}]{hwang2023strongly}%
  \BibitemOpen
  \bibfield  {author} {\bibinfo {author} {\bibfnamefont {Y.}~\bibnamefont {Hwang}}, \bibinfo {author} {\bibfnamefont {J.}~\bibnamefont {Puebla}}, \bibinfo {author} {\bibfnamefont {K.}~\bibnamefont {Kondou}}, \bibinfo {author} {\bibfnamefont {C.}~\bibnamefont {Gonzalez-Ballestero}}, \bibinfo {author} {\bibfnamefont {H.}~\bibnamefont {Isshiki}}, \bibinfo {author} {\bibfnamefont {C.~S.}\ \bibnamefont {Munoz}}, \bibinfo {author} {\bibfnamefont {L.}~\bibnamefont {Liao}}, \bibinfo {author} {\bibfnamefont {F.}~\bibnamefont {Chen}}, \bibinfo {author} {\bibfnamefont {W.}~\bibnamefont {Luo}}, \bibinfo {author} {\bibfnamefont {S.}~\bibnamefont {Maekawa}},  \emph {et~al.},\ }\href@noop {} {\bibfield  {journal} {\bibinfo  {journal} {arXiv preprint arXiv:2309.12690}\ } (\bibinfo {year} {2023})},\ \Eprint {http://arxiv.org/abs/2309.12690} {arXiv:2309.12690} \BibitemShut {NoStop}%
\bibitem [{\citenamefont {Geilen}\ \emph {et~al.}(2022{\natexlab{b}})\citenamefont {Geilen}, \citenamefont {Verba}, \citenamefont {Nicoloiu}, \citenamefont {Narducci}, \citenamefont {Dinescu}, \citenamefont {Ender}, \citenamefont {Mohseni}, \citenamefont {Ciubotaru}, \citenamefont {Weiler}, \citenamefont {Müller}, \citenamefont {Hillebrands}, \citenamefont {Adelmann},\ and\ \citenamefont {Pirro}}]{geilen2022parametric}%
  \BibitemOpen
  \bibfield  {author} {\bibinfo {author} {\bibfnamefont {M.}~\bibnamefont {Geilen}}, \bibinfo {author} {\bibfnamefont {R.}~\bibnamefont {Verba}}, \bibinfo {author} {\bibfnamefont {A.}~\bibnamefont {Nicoloiu}}, \bibinfo {author} {\bibfnamefont {D.}~\bibnamefont {Narducci}}, \bibinfo {author} {\bibfnamefont {A.}~\bibnamefont {Dinescu}}, \bibinfo {author} {\bibfnamefont {M.}~\bibnamefont {Ender}}, \bibinfo {author} {\bibfnamefont {M.}~\bibnamefont {Mohseni}}, \bibinfo {author} {\bibfnamefont {F.}~\bibnamefont {Ciubotaru}}, \bibinfo {author} {\bibfnamefont {M.}~\bibnamefont {Weiler}}, \bibinfo {author} {\bibfnamefont {A.}~\bibnamefont {Müller}}, \bibinfo {author} {\bibfnamefont {B.}~\bibnamefont {Hillebrands}}, \bibinfo {author} {\bibfnamefont {C.}~\bibnamefont {Adelmann}}, \ and\ \bibinfo {author} {\bibfnamefont {P.}~\bibnamefont {Pirro}},\ }\href@noop {} {\enquote {\bibinfo {title} {Parametric excitation and instabilities of spin waves driven by surface acoustic waves},}\ } (\bibinfo {year}
  {2022}{\natexlab{b}}),\ \Eprint {http://arxiv.org/abs/2201.04033} {arXiv:2201.04033} \BibitemShut {NoStop}%
\bibitem [{\citenamefont {Shah}\ \emph {et~al.}(2023)\citenamefont {Shah}, \citenamefont {Bas}, \citenamefont {Hamadeh}, \citenamefont {Wolf}, \citenamefont {Franson}, \citenamefont {Newburger}, \citenamefont {Pirro}, \citenamefont {Weiler},\ and\ \citenamefont {Page}}]{shah2023symmetry}%
  \BibitemOpen
  \bibfield  {author} {\bibinfo {author} {\bibfnamefont {P.~J.}\ \bibnamefont {Shah}}, \bibinfo {author} {\bibfnamefont {D.~A.}\ \bibnamefont {Bas}}, \bibinfo {author} {\bibfnamefont {A.}~\bibnamefont {Hamadeh}}, \bibinfo {author} {\bibfnamefont {M.}~\bibnamefont {Wolf}}, \bibinfo {author} {\bibfnamefont {A.}~\bibnamefont {Franson}}, \bibinfo {author} {\bibfnamefont {M.}~\bibnamefont {Newburger}}, \bibinfo {author} {\bibfnamefont {P.}~\bibnamefont {Pirro}}, \bibinfo {author} {\bibfnamefont {M.}~\bibnamefont {Weiler}}, \ and\ \bibinfo {author} {\bibfnamefont {M.~R.}\ \bibnamefont {Page}},\ }\href@noop {} {\enquote {\bibinfo {title} {Symmetry and nonlinearity of spin wave resonance excited by focused surface acoustic waves},}\ } (\bibinfo {year} {2023}),\ \Eprint {http://arxiv.org/abs/2305.06259} {arXiv:2305.06259} \BibitemShut {NoStop}%
\bibitem [{\citenamefont {Abrahams}\ and\ \citenamefont {Bernstein}(1969)}]{Abrahams1969Remeasurement}%
  \BibitemOpen
  \bibfield  {author} {\bibinfo {author} {\bibfnamefont {S.~C.}\ \bibnamefont {Abrahams}}\ and\ \bibinfo {author} {\bibfnamefont {J.~L.}\ \bibnamefont {Bernstein}},\ }\href {\doibase 10.1107/S0567740869003876} {\bibfield  {journal} {\bibinfo  {journal} {Acta Crystallographica Section B}\ }\textbf {\bibinfo {volume} {25}},\ \bibinfo {pages} {1233} (\bibinfo {year} {1969})}\BibitemShut {NoStop}%
\bibitem [{\citenamefont {Gurjar}\ \emph {et~al.}(2019)\citenamefont {Gurjar}, \citenamefont {Sharma}, \citenamefont {Patnaik},\ and\ \citenamefont {Kuanr}}]{Gurjar2019Structural}%
  \BibitemOpen
  \bibfield  {author} {\bibinfo {author} {\bibfnamefont {G.}~\bibnamefont {Gurjar}}, \bibinfo {author} {\bibfnamefont {V.}~\bibnamefont {Sharma}}, \bibinfo {author} {\bibfnamefont {S.}~\bibnamefont {Patnaik}}, \ and\ \bibinfo {author} {\bibfnamefont {B.~K.}\ \bibnamefont {Kuanr}},\ }\href {\doibase 10.1063/1.5113162} {\bibfield  {journal} {\bibinfo  {journal} {AIP Conference Proceedings}\ }\textbf {\bibinfo {volume} {2115}},\ \bibinfo {pages} {030323} (\bibinfo {year} {2019})}\BibitemShut {NoStop}%
\bibitem [{\citenamefont {Bachari}\ \emph {et~al.}(1999)\citenamefont {Bachari}, \citenamefont {Baud}, \citenamefont {{Ben Amor}},\ and\ \citenamefont {Jacquet}}]{bachari1999structural}%
  \BibitemOpen
  \bibfield  {author} {\bibinfo {author} {\bibfnamefont {E.}~\bibnamefont {Bachari}}, \bibinfo {author} {\bibfnamefont {G.}~\bibnamefont {Baud}}, \bibinfo {author} {\bibfnamefont {S.}~\bibnamefont {{Ben Amor}}}, \ and\ \bibinfo {author} {\bibfnamefont {M.}~\bibnamefont {Jacquet}},\ }\href {\doibase https://doi.org/10.1016/S0040-6090(99)00060-7} {\bibfield  {journal} {\bibinfo  {journal} {Thin Solid Films}\ }\textbf {\bibinfo {volume} {348}},\ \bibinfo {pages} {165} (\bibinfo {year} {1999})}\BibitemShut {NoStop}%
\bibitem [{\citenamefont {Lin}\ and\ \citenamefont {Huang}(2004)}]{lin2004effect}%
  \BibitemOpen
  \bibfield  {author} {\bibinfo {author} {\bibfnamefont {S.-S.}\ \bibnamefont {Lin}}\ and\ \bibinfo {author} {\bibfnamefont {J.-L.}\ \bibnamefont {Huang}},\ }\href {\doibase https://doi.org/10.1016/j.surfcoat.2003.11.014} {\bibfield  {journal} {\bibinfo  {journal} {Surface and Coatings Technology}\ }\textbf {\bibinfo {volume} {185}},\ \bibinfo {pages} {222} (\bibinfo {year} {2004})}\BibitemShut {NoStop}%
\bibitem [{\citenamefont {Howe}\ \emph {et~al.}(2015)\citenamefont {Howe}, \citenamefont {Emori}, \citenamefont {Jeon}, \citenamefont {Oxholm}, \citenamefont {Jones}, \citenamefont {Mahalingam}, \citenamefont {Zhuang}, \citenamefont {Sun},\ and\ \citenamefont {Brown}}]{howe2015pseudomorphic}%
  \BibitemOpen
  \bibfield  {author} {\bibinfo {author} {\bibfnamefont {B.~M.}\ \bibnamefont {Howe}}, \bibinfo {author} {\bibfnamefont {S.}~\bibnamefont {Emori}}, \bibinfo {author} {\bibfnamefont {H.-M.}\ \bibnamefont {Jeon}}, \bibinfo {author} {\bibfnamefont {T.~M.}\ \bibnamefont {Oxholm}}, \bibinfo {author} {\bibfnamefont {J.~G.}\ \bibnamefont {Jones}}, \bibinfo {author} {\bibfnamefont {K.}~\bibnamefont {Mahalingam}}, \bibinfo {author} {\bibfnamefont {Y.}~\bibnamefont {Zhuang}}, \bibinfo {author} {\bibfnamefont {N.~X.}\ \bibnamefont {Sun}}, \ and\ \bibinfo {author} {\bibfnamefont {G.~J.}\ \bibnamefont {Brown}},\ }\href {\doibase 10.1109/LMAG.2015.2449260} {\bibfield  {journal} {\bibinfo  {journal} {IEEE Magnetics Letters}\ }\textbf {\bibinfo {volume} {6}},\ \bibinfo {pages} {1} (\bibinfo {year} {2015})}\BibitemShut {NoStop}%
\bibitem [{\citenamefont {Foster}\ \emph {et~al.}(1968)\citenamefont {Foster}, \citenamefont {Coquin}, \citenamefont {Rozgonyi},\ and\ \citenamefont {Vannatta}}]{foster1968cadmium}%
  \BibitemOpen
  \bibfield  {author} {\bibinfo {author} {\bibfnamefont {N.}~\bibnamefont {Foster}}, \bibinfo {author} {\bibfnamefont {G.}~\bibnamefont {Coquin}}, \bibinfo {author} {\bibfnamefont {G.}~\bibnamefont {Rozgonyi}}, \ and\ \bibinfo {author} {\bibfnamefont {F.}~\bibnamefont {Vannatta}},\ }\href {\doibase 10.1109/T-SU.1968.29443} {\bibfield  {journal} {\bibinfo  {journal} {IEEE Transactions on Sonics and Ultrasonics}\ }\textbf {\bibinfo {volume} {15}},\ \bibinfo {pages} {28} (\bibinfo {year} {1968})}\BibitemShut {NoStop}%
\bibitem [{\citenamefont {Yanagitani}\ \emph {et~al.}(2011)\citenamefont {Yanagitani}, \citenamefont {Morisato}, \citenamefont {Takayanagi}, \citenamefont {Matsukawa},\ and\ \citenamefont {Watanabe}}]{Yanagitani2011cAxisZZ}%
  \BibitemOpen
  \bibfield  {author} {\bibinfo {author} {\bibfnamefont {T.}~\bibnamefont {Yanagitani}}, \bibinfo {author} {\bibfnamefont {N.}~\bibnamefont {Morisato}}, \bibinfo {author} {\bibfnamefont {S.}~\bibnamefont {Takayanagi}}, \bibinfo {author} {\bibfnamefont {M.}~\bibnamefont {Matsukawa}}, \ and\ \bibinfo {author} {\bibfnamefont {Y.}~\bibnamefont {Watanabe}},\ }\href {https://api.semanticscholar.org/CorpusID:22175434} {\bibfield  {journal} {\bibinfo  {journal} {IEEE Transactions on Ultrasonics, Ferroelectrics and Frequency Control}\ }\textbf {\bibinfo {volume} {58}} (\bibinfo {year} {2011})}\BibitemShut {NoStop}%
\bibitem [{\citenamefont {Laude}\ and\ \citenamefont {Ballandras}(2003)}]{laude2003slowness}%
  \BibitemOpen
  \bibfield  {author} {\bibinfo {author} {\bibfnamefont {V.}~\bibnamefont {Laude}}\ and\ \bibinfo {author} {\bibfnamefont {S.}~\bibnamefont {Ballandras}},\ }\href {\doibase 10.1063/1.1582237} {\bibfield  {journal} {\bibinfo  {journal} {Journal of Applied Physics}\ }\textbf {\bibinfo {volume} {94}},\ \bibinfo {pages} {1235} (\bibinfo {year} {2003})}\BibitemShut {NoStop}%
\bibitem [{\citenamefont {O'Rorke}\ \emph {et~al.}(2020)\citenamefont {O'Rorke}, \citenamefont {Winkler}, \citenamefont {Collins},\ and\ \citenamefont {Ai}}]{o2020slowness}%
  \BibitemOpen
  \bibfield  {author} {\bibinfo {author} {\bibfnamefont {R.}~\bibnamefont {O'Rorke}}, \bibinfo {author} {\bibfnamefont {A.}~\bibnamefont {Winkler}}, \bibinfo {author} {\bibfnamefont {D.}~\bibnamefont {Collins}}, \ and\ \bibinfo {author} {\bibfnamefont {Y.}~\bibnamefont {Ai}},\ }\href {\doibase 10.1039/C9RA10452F} {\bibfield  {journal} {\bibinfo  {journal} {RSC advances}\ }\textbf {\bibinfo {volume} {10}},\ \bibinfo {pages} {11582} (\bibinfo {year} {2020})}\BibitemShut {NoStop}%
\bibitem [{\citenamefont {Maupin}(2007)}]{maupin2007introduction}%
  \BibitemOpen
  \bibfield  {author} {\bibinfo {author} {\bibfnamefont {V.}~\bibnamefont {Maupin}},\ }in\ \href {\doibase https://doi.org/10.1016/S0065-2687(06)48002-X} {\emph {\bibinfo {booktitle} {Advances in Wave Propagation in Heterogenous Earth}}},\ \bibinfo {series} {Advances in Geophysics}, Vol.~\bibinfo {volume} {48},\ \bibinfo {editor} {edited by\ \bibinfo {editor} {\bibfnamefont {R.-S.}\ \bibnamefont {Wu}}, \bibinfo {editor} {\bibfnamefont {V.}~\bibnamefont {Maupin}}, \ and\ \bibinfo {editor} {\bibfnamefont {R.}~\bibnamefont {Dmowska}}}\ (\bibinfo  {publisher} {Elsevier},\ \bibinfo {year} {2007})\ pp.\ \bibinfo {pages} {127--155}\BibitemShut {NoStop}%
\end{thebibliography}%



\end{document}

\title{SUPPLEMENTAL MATERIAL \protect\\
Generation of gigahertz frequency surface acoustic waves in YIG/ZnO heterostructures}%

\author{Finlay Ryburn}
\email{finlay.ryburn@physics.ox.ac.uk}
\affiliation{Clarendon Laboratory, Department of Physics, University of Oxford, Parks Road, Oxford, OX1\,3PU, United Kingdom}
\author{Kevin Künstle}
\email{kuenstle@rptu.de}
\affiliation{Fachbereich Physik and Landesforschungszentrum OPTIMAS, Rheinland-Pfälzische Technische Universität Kaiserslautern-Landau, 67663 Kaiserslautern, Germany}
\author{Yangzhan Zhang}
\affiliation{Clarendon Laboratory, Department of Physics, University of Oxford, Parks Road, Oxford, OX1\,3PU, United Kingdom}
\author{Yannik Kunz}
\affiliation{Fachbereich Physik and Landesforschungszentrum OPTIMAS, Rheinland-Pfälzische Technische Universität Kaiserslautern-Landau, 67663 Kaiserslautern, Germany}
\author{Timmy Reimann}
\affiliation{INNOVENT e.V. Technologieentwicklung, 07745 Jena, Germany}
\author{Morris Lindner}
\affiliation{INNOVENT e.V. Technologieentwicklung, 07745 Jena, Germany}
\author{Carsten Dubs}
\affiliation{INNOVENT e.V. Technologieentwicklung, 07745 Jena, Germany}
\author{John F. Gregg}
\affiliation{Clarendon Laboratory, Department of Physics, University of Oxford, Parks Road, Oxford, OX1\,3PU, United Kingdom}
\author{Mathias Weiler}
\affiliation{Fachbereich Physik and Landesforschungszentrum OPTIMAS, Rheinland-Pfälzische Technische Universität Kaiserslautern-Landau, 67663 Kaiserslautern, Germany}

\date{\today}%

\maketitle

\clearpage
\newpage

\renewcommand{\figurename}{Fig.}
\renewcommand{\thefigure}{S\arabic{figure}}

\renewcommand{\tablename}{TABLE}
\renewcommand{\thetable}{S\arabic{table}}

\section*{[S.1] Sample preparation}

To investigate the surface acoustic wave (SAW) properties in the YIG/ZnO heterostructures, we fabricated interdigital transducers (IDTs) via electron beam lithography (EBL) on a YIG/GGG bilayer, which were subsequently covered with a piezoelectric ZnO film. Chemical etching was employed to remove the ZnO from the contact pads of the IDTs. This section provides an in-depth description of the fabrication process.

The 103nm and 55\,nm thick YIG thin films employed in this study are grown in the (111) direction on 500\,$\mathrm{\mu}$m GGG substrates via liquid phase epitaxy. Prior to the structuring process, the chips were cleaned with acetone and isopropanol in an ultrasonic bath at 35\,kHz. Subsequently, a double resist layer consisting of PMMA 600K 4\% and PMMA 950K 2\% was spin-coated onto the YIG. The different resists result in an undercut after the electron beam lithography, needed for a successful lift-off process. An additional layer of Espacer300Z is applied to the PMMA surface in order to prevent surface charge-up effects. To achieve SAW-wavelengths on the $\mathrm{\mu}$m scale, a RAITH VOYAGER EBL system was used to fabricate the IDT fingers, which have a width and spacing of 700\,nm. Following resist development, small PMMA residues that might remain in the mask grooves are removed via plasma ashing. Subsequently, the sample was covered with a 10\,nm titanium, 80\,nm gold stack using electron beam evaporation. After the electron beam evaporation of the two metals, the resist lift-off takes place and only the metal in the mask grooves, that is in contact with YIG, remains on the sample surface. 

\begin{figure}
\centering
\vspace{0pt}
\includegraphics[width=0.49\textwidth]{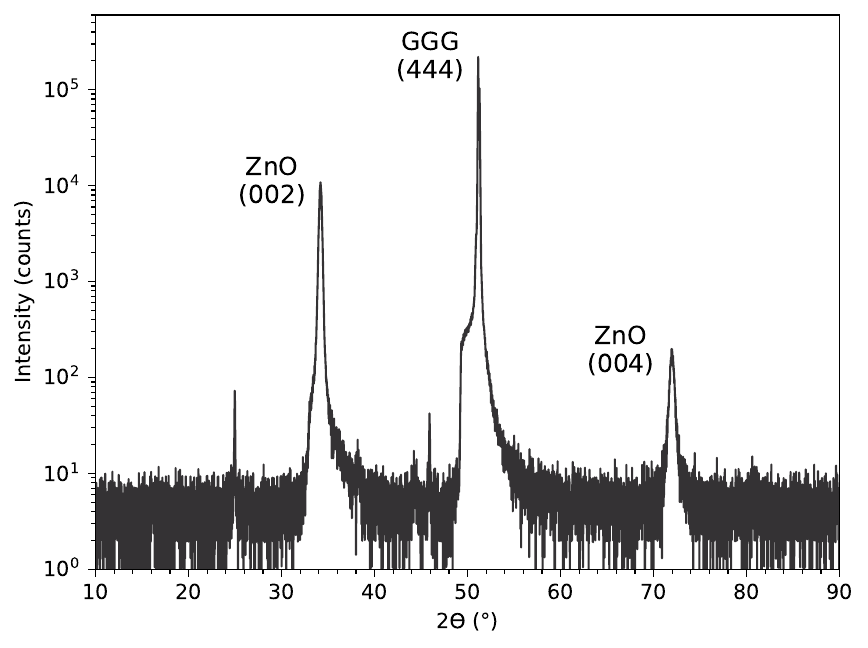}
\caption{Sample characterisation, $2\theta / \theta$ x-ray diffraction data for Sample 1~\cite{bachari1999structural, lin2004effect, howe2015pseudomorphic}.}
\label{fig:XRD}
\vspace{0pt}
\end{figure}

The ensuing structures are then covered with a piezoelectric zinc oxide film which was grown on the whole chip by radio frequency magnetron sputtering. The base pressure in the sputtering chamber was 10$^{-5}$\,mbar, and the sputtering gas pressure approximately $\num{4e-3}$\,mbar. To achieve the right stoichiometry, a sputtering gas mixture of argon and oxygen with a 2:1 ratio was used. The sample was first pre-sputtered at 30\,W for 5 minutes, before sputtering was performed at power levels of 140\,W and 100\,W, resulting in ZnO films with thicknesses of 890$\pm$25\,nm for Sample 1 and 830$\pm$25\,nm for Sample 2, as measured using a profilometer. The relatively large errors in the thickness result from a combination of surface defects and non-uniformity in the thickness due to the sputtering process, a single 90\,mm diameter ZnO target was used; to increase the accuracy we averaged over 5 measurements close to the excited IDTs. The insulating ZnO now also covers the contacting pads of the IDT structures and needs to be removed to allow direct electrical contact. This is accomplished by creating an etch mask consisting of AZ5214E photoresist using a Microtech LW405D LaserWriter. Following resist development, the sample was immersed in a 32\% HCl solution removing the ZnO from the contact pads. XRD $2\theta / \theta$ characterization of the samples was later carried out with a Rigaku SMARTLab diffractometer, the results for Sample 1 are shown in Fig.~\ref{fig:XRD}, to verify the crystalline orientation. The peak at approximately \ang{34} indicates the successful deposition of ZnO with an out-of-plane c-axis~\cite{bachari1999structural, lin2004effect}.

\section*{[S.2] Material parameters}

The material properties used in the analytical calculation are given in Tab.~\ref{table:materialproperties}.
\begin{table}[ht]
\caption{The material properties for the different layers used in the analytical calculations. Density is given in kgm$^{-3}$, the elastic constants in $10^{9}$Nm$^{-2}$, the piezoelectric constants in Cm$^{-2}$, and the permittivities in $10^{-11}$Fm$^{-1}$. The relevant references are ZnO~\cite{zhang_2022_bulk}, YIG~\cite{clark1961elastic, hasan2017yttrium}, and GGG~\cite{graham1970elastic, connelly2021complex}.} % title of Table
\centering % used for centering table
\begin{tabular}{c c c c c c c c c c c c c } % centered columns (4 columns)
\hline\hline %inserts double horizontal lines
Material & Class & $\rho$ & $c_{11}$ & $c_{12}$ & $c_{13}$ & $c_{33}$ & $c_{44}$ & $e_{x5}$ & $e_{z1}$ & $e_{z3}$ & $\epsilon_{xx}$ & $\epsilon_{zz}$\\ [0.5ex] % inserts table
\hline % inserts single horizontal line
ZnO & 6mm & 5700 & 209 & 121.1 & 105.1 & 210.9 & 42.5 & -0.48 & -0.57 & 1.32 & 8.55 & 10.2\\ % inserting body of the table
YIG & m$\bar{3}$m & 5170 & 269 &  107.7 &  107.7 & 269 & 76.4 & 0 & 0 & 0 & 4.3 & 4.3\\
GGG & m$\bar{3}$m & 7094 & 285.7 & 114.9 & 114.9 & 285.7 & 90.2 & 0 & 0 & 0 & 12.1 & 12.1\\
`Dead layer' & N/A & 5700 & 204.7 & 112.3 & 113.3 & 205.6 & 45.4 & 0 & 0 & 0 & 9.2 & 9.0\\[1ex] % [1ex] adds vertical space
\hline %inserts single line
\end{tabular}
\label{table:materialproperties} % is used to refer this table in the text
\end{table}

\section*{[S.3] Coupling coefficients, displacements, and strains}

\begin{figure}[!ht]
\centering
\vspace{0pt}
\includegraphics[width=0.49\textwidth]{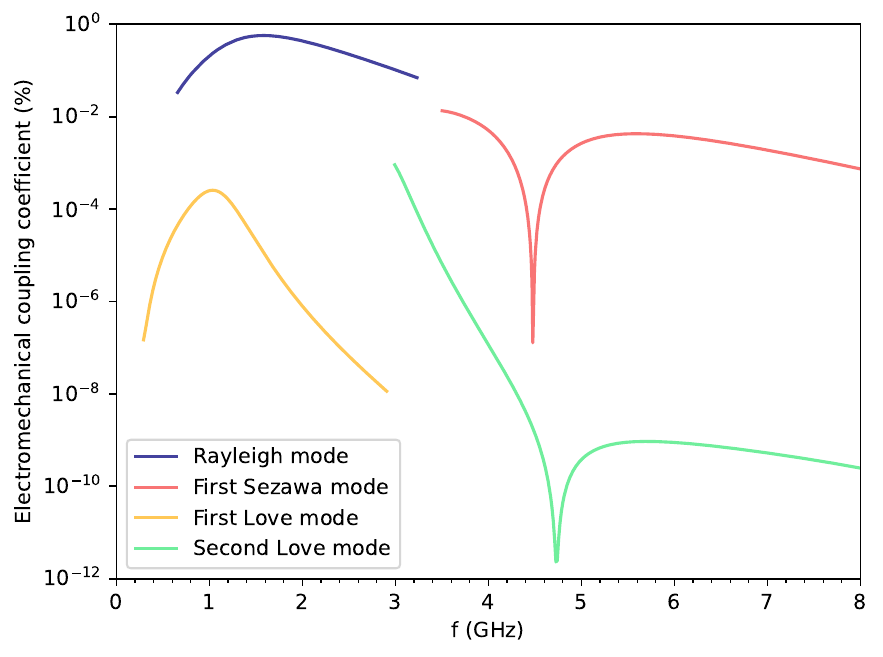}
\caption{The electromechanical coupling coefficients of the four lowest order modes calculated for Sample 1 as a function of frequency.}
\label{fig:coupling}
\vspace{0pt}
\end{figure}

The calculated electromechanical coupling coefficients for Sample 1 can be seen in Fig.~\ref{fig:coupling}. As discussed in Section IV of the main text, the Love-like modes have coupling coefficients one to two orders of magnitude lower than those of the Rayleigh-like modes, hence our decision to neglect this class of solutions.

\begin{figure}[!ht]
\centering
\vspace{0pt}
\includegraphics[width=0.49\textwidth]{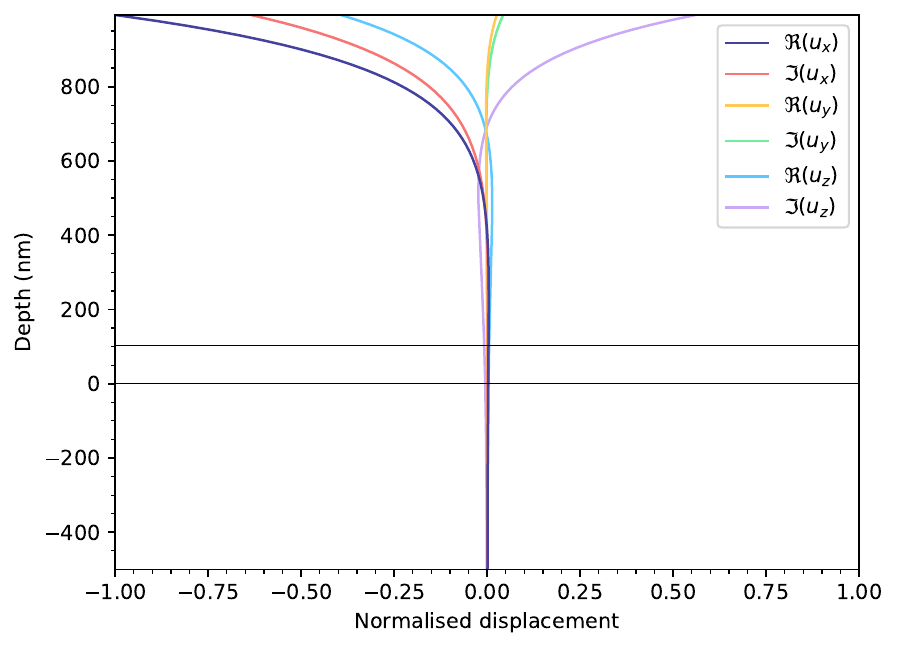}
\caption{The real and imaginary components of the normalised displacements as a function of depth in Sample 1 for the Rayleigh-like mode at 2.9\,GHz. The two horizontal black lines indicate the position of the YIG layer, with depth zero at the GGG/YIG interface.}
\label{fig:Displacements}
\vspace{0pt}
\end{figure}

The normalised displacements, Fig.~\ref{fig:Displacements}, and normalised strains at the surface, Tab.~\ref{table:strainRayleighSurface}, for Sample 1 are calculated at fixed frequencies. We see that the displacements decay exponentially and that the dominant strain contributions are $S_{xx}$, $S_{zz}$, and $S_{xz}$, with the usual phase differences, 
as expected for typical Rayleigh mode solutions~\cite{FARNELL197235, kuss2020nonreciprocal}. We note that due to the anisotropic layered materials, the particle motion is not purely restricted to the $\boldsymbol{x_1}$, $\boldsymbol{x_3}$ plane as in the isotropic case~\cite{FARNELL197235}; there are also small displacements in the $\boldsymbol{x_2}$ direction.

\begin{table}[ht]
\caption{The normalised real, imaginary, and total magnitude of the strain components for the Rayleigh-like mode at 2.9\,GHz and the first Sezawa-like mode at 5.65\,GHz calculated at the surface for Sample 1.} % title of Table
\centering % used for centering table
\begin{tabular}{c c c c c c c} % centered columns (4 columns)
\hline\hline %inserts double horizontal lines
 & \multicolumn{3}{c}{Rayleigh 2.9\,GHz} & \multicolumn{3}{c}{Sezawa 5.65\,GHz} \\
\hline
$S_{ij}$ & $\mathfrak{R} (S_{ij})$ & $\mathfrak{I}(S_{ij})$ & $|S_{ij}|$ & $\mathfrak{R} (S_{ij})$ & $\mathfrak{I}(S_{ij})$ & $|S_{ij}|$\\ [0.5ex] % inserts table
%heading
\hline % inserts single horizontal line
$S_{xx}$ & 0.65 & -1.00 & 1.00 & 1.00 & -0.50 & 1.00\\ % inserting body of the table
$S_{yy}$ & 0.00 & 0.00 & 0.00 & 0.00 & 0.00 & 0.00\\
$S_{zz}$ & -0.81 & -0.52 & 0.73 & -0.81 & 0.38 & 0.80\\
$S_{xy}$ & -0.02 & 0.01 & 0.02 & -0.07 & 0.00 & 0.06 \\
$S_{xz}$ & -0.81 & -0.52 & 0.81 & -0.43 & -0.88 & 0.88 \\ 
$S_{yz}$ & 0.02 & 0.03 & 0.03 & 0.01 & 0.07 & 0.06 \\ [1ex] % [1ex] adds vertical space
\hline %inserts single line
\end{tabular}
\label{table:strainRayleighSurface} % is used to refer this table in the text
\end{table}

\section*{[S.4] Rotating crystallographic axes}

When considering the `dead layer', and indeed the orientation of any of the crystal layers, it is necessary to rotate the crystallographic axes. Practically, this involves the rotation of the permittivity, piezoelectric, and elastic tensors used in the model. We consider an arbitrary rotation $\alpha$ about $\boldsymbol{x_1}$, followed by $\beta$ about $\boldsymbol{x_2}$, and finally $\gamma$ about $\boldsymbol{x_3}$. This gives the 3x3 rotation matrix
\begin{equation}
\begin{aligned}
a=
\begin{pmatrix}
\cos \gamma & -\sin \gamma  & 0\\
\sin \gamma & \cos \gamma & 0 \\
0 & 0 & 1
\end{pmatrix}\begin{pmatrix}
\cos \beta & 0 & \sin \beta\\
 0 & 1 & 0 \\
-\sin \beta & 0 & \cos \beta
\end{pmatrix}\begin{pmatrix}
1 & 0 & 0 \\
0 & \cos \alpha & -\sin \alpha\\
0 & \sin \alpha & \cos \alpha 
\end{pmatrix}
=\begin{pmatrix}
a_{11} & a_{12} & a_{13}\\
a_{21} & a_{22} & a_{23} \\
a_{31} & a_{32} & a_{33}
\end{pmatrix}.
\end{aligned}
\end{equation}
The corresponding 6x6 rotation matrix is given by~\cite{zhang_2022_bulk}
\begingroup
\renewcommand*{\arraystretch}{1.2}
\begin{equation}
\begin{aligned}
M=
\begin{pmatrix}
a_{11}^2& a_{12}^2& a_{13}^2& 2 a_{12} a_{13}& 2 a_{13} a_{11}& 2 a_{11} a_{12}\\
a_{21}^2& a_{22}^2& a_{23}^2& 2 a_{22} a_{23}& 2 a_{23} a_{21}& 2 a_{21} a_{22}\\
a_{31}^2& a_{32}^2& a_{33}^2& 2 a_{32} a_{33}& 2 a_{33} a_{31}& 2 a_{31} a_{32}\\
a_{21} a_{31}& a_{22} a_{32}& a_{23} a_{33}& a_{22} a_{33}+a_{23} a_{32}& a_{21} a_{33}+a_{23} a_{31}& a_{22} a_{31}+a_{21} a_{32}\\
a_{31} a_{11}& a_{32} a_{12}& a_{33} a_{13}& a_{12} a_{33}+a_{13} a_{32}& a_{13} a_{31}+a_{11} a_{33}& a_{11} a_{32}+a_{12} a_{31}\\
a_{11} a_{21}& a_{12} a_{22}& a_{13} a_{23}& a_{12} a_{23}+a_{13} a_{22}& a_{13} a_{21}+a_{11} a_{23}& a_{11} a_{22}+a_{12} a_{21}
\end{pmatrix}
\end{aligned}
\end{equation}
\endgroup
and the tensors are transformed according to
\begin{equation}
    \epsilon ' = a \epsilon a^T, \hspace{4pt} e' = a e M^T, \hspace{4pt} c' = M c M^T. \label{eq:rotation}
\end{equation}

\section*{[S.5] The `dead layer'}

As discussed, we model the `dead layer', where XRD data indicates there is no alignment of the c-axis~\cite{fung2015phonon}, by considering a randomly oriented c-axis. To calculate the material properties of an arbitrarily aligned c-axis crystallite in this layer, we apply the transformations to the permittivity and elastic tensors given in Eq.(\ref{eq:rotation}) for arbitrary angles and then find the average properties by integrating over all possible angles~\cite{den1999relation}
\begin{equation}
    \epsilon_{\mathrm{dead}} ' =  \tplint{2\pi}{0}{2\pi}{0}{2\pi}{0}\epsilon'(\alpha, \beta, \gamma) \;d\alpha\,d\beta\,d\gamma, \hspace{6pt} c_{\mathrm{dead}}' = \tplint{2\pi}{0}{2\pi}{0}{2\pi}{0}c'(\alpha, \beta, \gamma) \;d\alpha\,d\beta\,d\gamma.
\end{equation}
These transformations lead to relatively small changes in the material properties and tend to symmetrise the matrices as expected, see Tab.~\ref{table:materialproperties}. Finally, we assume any action of the electric dipoles to cancel globally and thus set the piezoelectric tensor to zero, and assume the density of the ZnO remains unchanged. 

\begin{figure*}[!ht]
\subfloat[\label{fig:deadlayervelocity}]{%
  \includegraphics[width=.32\textwidth]{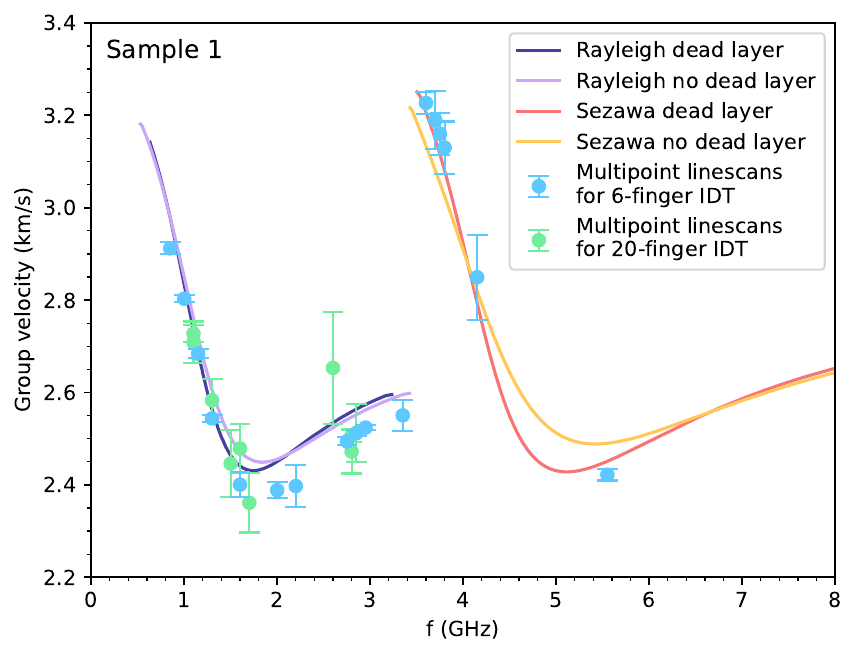}%
}\hfill
\subfloat[\label{fig:deadlayervelocitySample2}]{%
  \includegraphics[width=.32\textwidth]{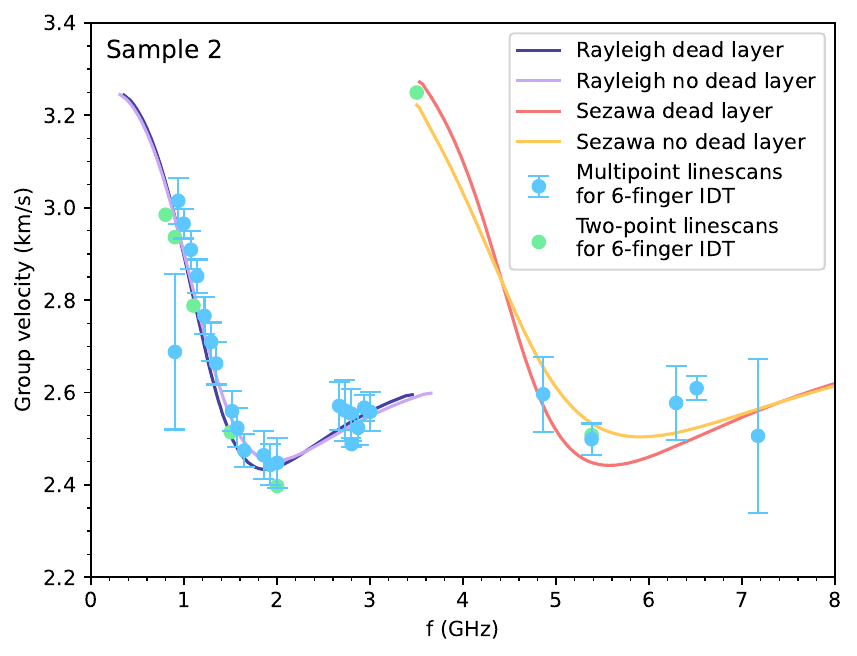}%
}\hfill
\subfloat[\label{fig:deadlayercoupling}]{%
  \includegraphics[width=.32\textwidth]{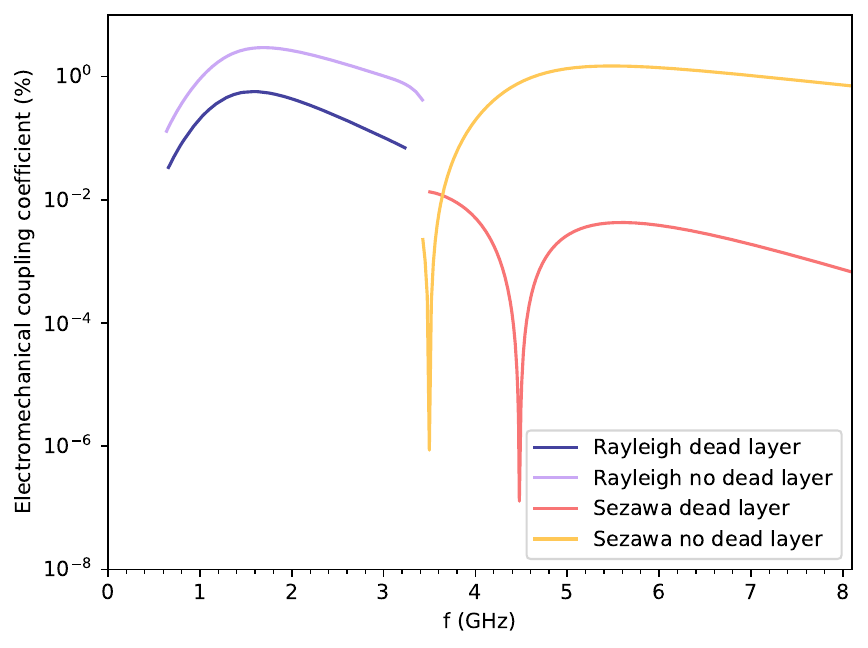}%
}
\caption{The experimental group velocity data in light blue and green for Sample 1 (a) and Sample 2 (b) plotted against the analytical solutions with (dark blue and red) and without (lilac and yellow) the `dead layer' included in the calculation. (c) The electromechanical coupling constants calculated with (dark blue and red) and without (lilac and yellow) the `dead layer'. }
\end{figure*}

The effect on the group velocity for Samples 1 \& 2 on the addition of the `dead layer' can be seen in Fig.~\ref{fig:deadlayervelocity}. We see a fairly minimal change with a deepening of the dip, which seems to slightly improve the fit to the experimental data. Given the nature of the model used, it is likely that the real group velocity curve lies somewhere between these curves as the model takes the worst-case scenario where the `dead layer' has no coherence in the alignment of the electric dipoles. This is also likely true of the coupling coefficients, Fig.~\ref{fig:deadlayercoupling}, where we see a drop in the expected coupling by one to two orders of magnitude. However, given the results of the detailed study in~\cite{fung2015phonon} we consider it more accurate to include this approximate model rather than to neglect the `dead layer' entirely. 

\section*{[S.6] Dependence on c-axis angle and YIG orientation}

\begin{figure*}[ht]
\subfloat[\label{fig:slownessYIG}]{%
  \includegraphics[width=.32\textwidth]{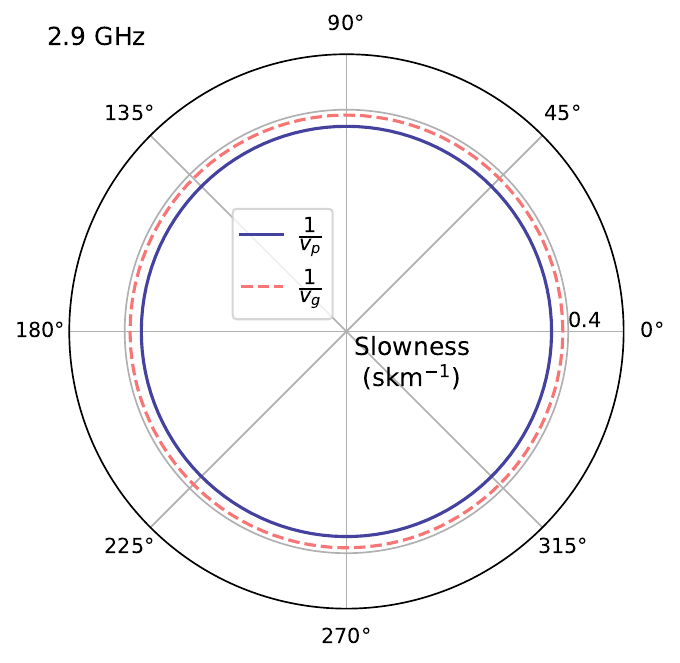}%
}\hfill
\subfloat[\label{fig:slownessRayleigh}]{%
  \includegraphics[width=.32\textwidth]{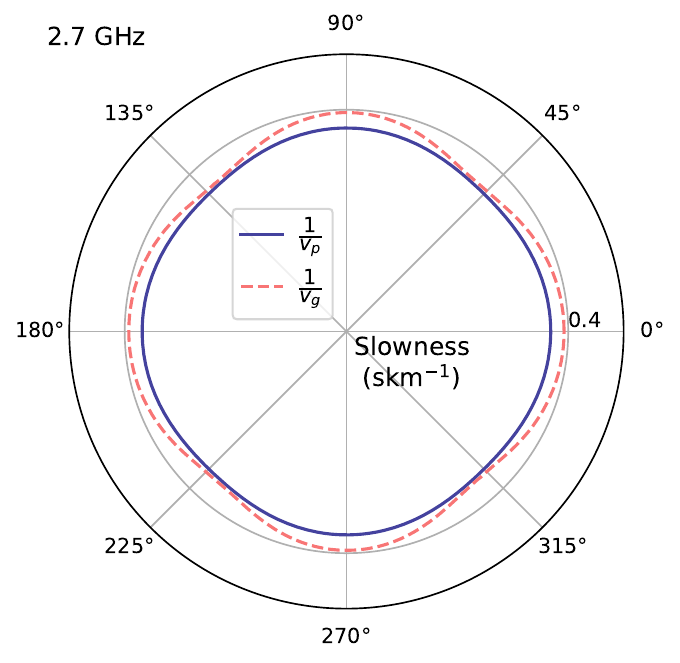}%
}\hfill
\subfloat[\label{fig:slownessSezawa}]{%
  \includegraphics[width=.32\textwidth]{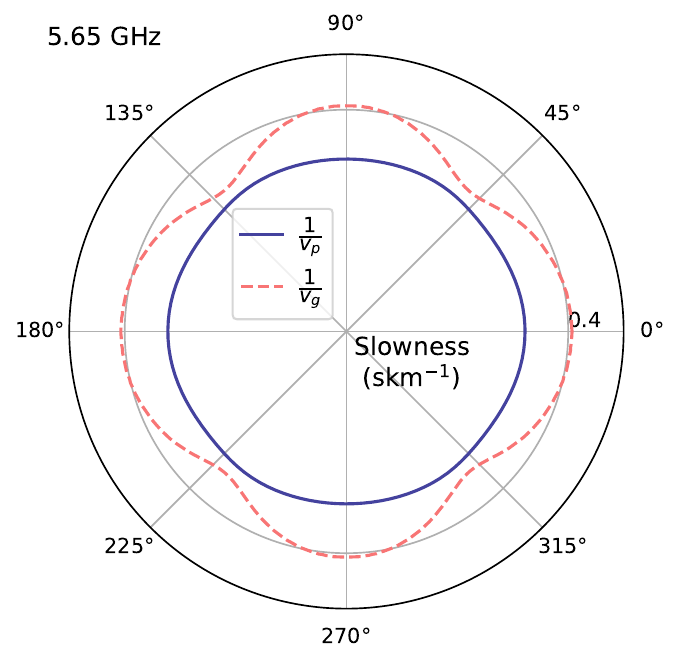}%
}
\caption{Slowness curves for the rotation of different crystallographic axes at fixed frequencies. (a) Slowness as a function of the in-plane GGG/YIG crystallographic axis orientation angle at a fixed frequency of 2.9\,GHz for the Rayleigh-like mode. Rotation is about the $\boldsymbol{x_3}$ axis, and \ang{0} corresponds to the (1,1,-2) direction along $\boldsymbol{x_1}$. (b) Slowness as a function of the ZnO c-axis orientation angle at a fixed frequency of 2.7\,GHz for the Rayleigh-like mode. (c) Slowness as a function of the ZnO c-axis orientation angle at a fixed frequency of 5.65\,GHz for the Sezawa-like mode. For (b) and (c) rotation is about the $\boldsymbol{x_2}$ axis, and \ang{0} corresponds to an out-of-plane oriented c-axis.}
\end{figure*}

We also consider whether the in-plane orientation of the GGG/YIG crystallographic axes affects the SAW properties. To do this, we rotate the GGG and YIG tensors about the out-of-plane $\boldsymbol{x_3}$ axis and evaluate the dispersion relation at each angle. The results of this calculation at a fixed frequency of 2.9\,GHz can be seen in Fig.~\ref{fig:slownessYIG}. We plot the slowness curves, where slowness is defined as the inverse of the velocity (we consider both group and phase velocities). It can be seen the phase and group velocities of the SAW are isotropic with the rotation of this axis.

The perhaps more interesting property to vary is the ZnO piezoelectric c-axis orientation. This is commonly studied in the case of bulk acoustic wave transducers and has a significant effect on the electromechanical coupling coefficient. In the case of an out-of-plane c-axis the waves are purely longitudinal, whereas, with the c-axis aligned at $\sim$\ang{40} to the normal of the plane, the waves are purely shear~\cite{foster1968cadmium, Yanagitani2011cAxisZZ, zhang_2022_bulk}. Here we consider rotation about the $\boldsymbol{x_2}$ axis where \ang{0} corresponds to an out-of-plane oriented c-axis. The slowness curves can be seen in Fig.~\ref{fig:slownessRayleigh} for the Rayleigh-like mode at 2.7\,GHz, and Fig.~\ref{fig:slownessSezawa} for the Sezawa-like mode at 5.65\,GHz. Here we no longer see isotropic behaviour and have the waves travelling at maximum velocity when the c-axis is oriented at \ang{48}. This suggests changing the orientation angle of the c-axis to approximately \ang{48} may be sensible to avoid beam steering effects~\cite{laude2003slowness, o2020slowness}. Note, we consider 2.7\,GHz rather than 2.9\,GHz as the Rayleigh-like solutions at some intermediate angles do not exist for frequencies as high as 2.9\,GHz. 

\begin{figure*}[ht]
\subfloat[\label{fig:couplingRayleigh}]{%
  \includegraphics[width=.49\textwidth]{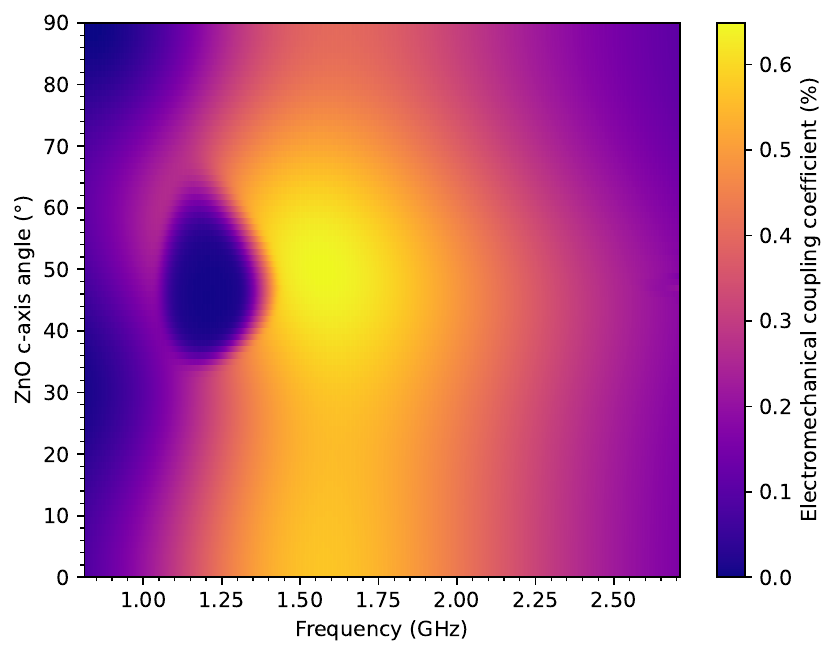}%
}\hfill
\subfloat[\label{fig:couplingSezawa}]{%
  \includegraphics[width=.49\textwidth]{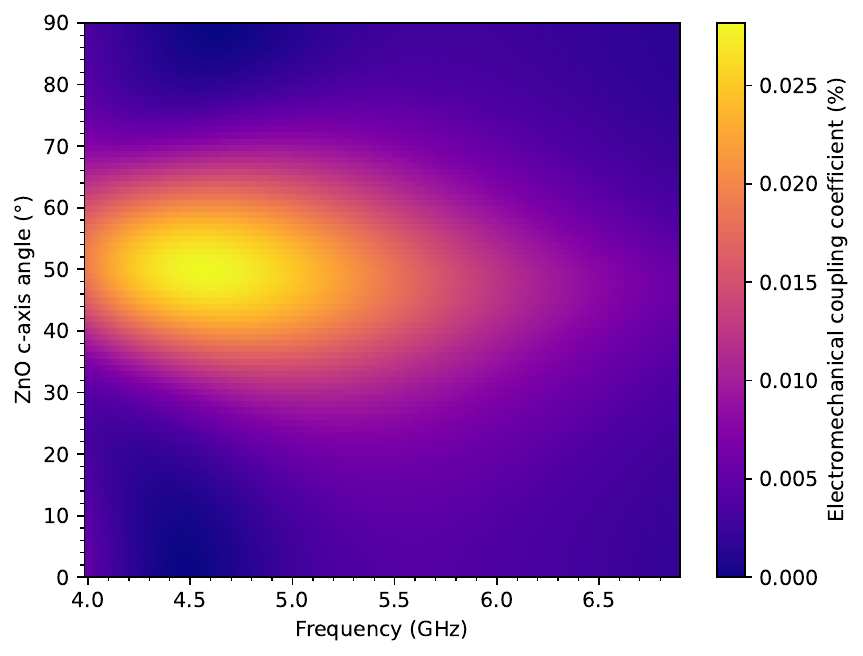}%
}
\caption{Electromechanical coupling constant as a function of ZnO c-axis angle and frequency for the Rayleigh-like mode (a) and Sezawa-like mode (b). Rotation is about the $x_2$ axis, and \ang{0} corresponds to an out-of-plane oriented c-axis.}
\end{figure*}

Finally, we evaluate the electromechanical coupling coefficients as a function of the ZnO c-axis orientation over a range of frequencies. The resulting coupling coefficients for the Rayleigh-like mode can be seen in Fig.~\ref{fig:couplingRayleigh} and for the Sezawa-like mode in Fig.~\ref{fig:couplingSezawa}. For both the Rayleigh and Sezawa-like modes, we observe maxima in the calculated coupling coefficients with a c-axis angle of approximately \ang{50}. This suggests $\sim$\,\ang{50} is the optimum c-axis angle for maximum velocity and SAW excitation efficiency. In the case of the Rayleigh-like mode, we observe an interesting minima in the intensity at around \ang{50} and 1.3\,GHz, this occurs due to mode mixing~\cite{maupin2007introduction, FARNELL197235} between the Rayleigh-like mode and the first Love-like mode as their phase velocities intersect. This demonstrates the importance of carrying out these calculations before choosing the IDT wavelength, as one could inadvertently design an IDT to work at such a minimum.  

Using these results, the ZnO c-axis angle and thickness can be tuned to optimise a certain mode at a given frequency to maximise the electromechanical coupling, thus producing the strongest SAW intensity for a given IDT wavelength. Alongside the potentially increased efficiency and reduced input power, this is useful if attempting to enhance a spin wave signal if studying magnetoelastic coupling.

\begin{figure}
\centering
\vspace{0pt}
\includegraphics[width=0.49\textwidth]{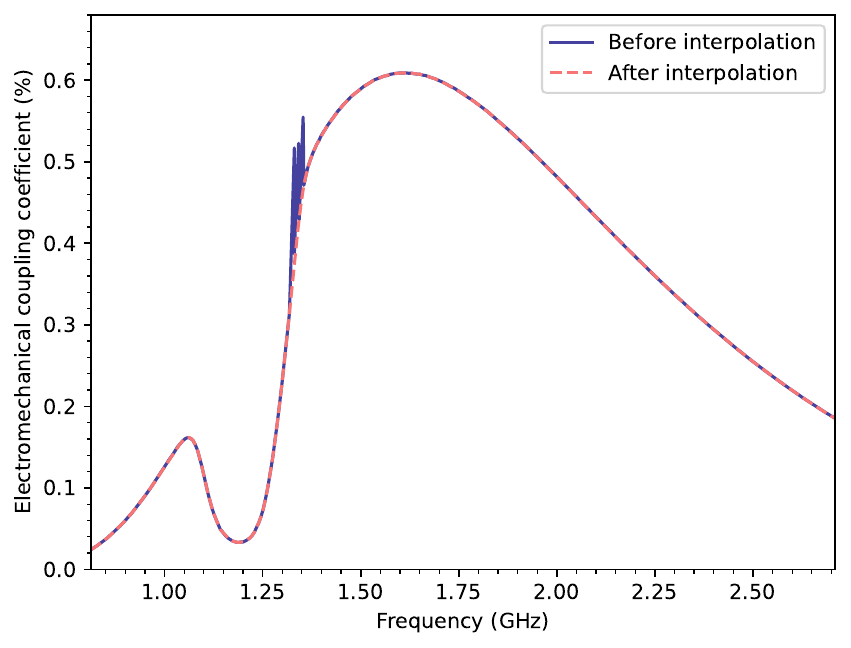}
\caption{The electromechanical coupling coefficient as a function of frequency with the ZnO c-axis angled at \ang{38} to the surface normal. The raw results of the calculation are shown in dark blue and the results after interpolation are shown in dashed red.}
\label{fig:interpolation}
\vspace{0pt}
\end{figure}

We note that for the data in Fig.~\ref{fig:couplingRayleigh}, a sharp discontinuity occurred where the calculation converged between the Rayleigh-like and Love-like solutions at different positions for the free and metallic phase velocities, before consistently converging on the Rayleigh-like solution as the phase velocities of the two modes stop mixing. This occurred over a small range, approximately 0.05\,GHz centred on $\sim$\,1.4\,GHz between \ang{37} and \ang{59}. We eliminated this by deleting the points over the discontinuity and interpolating between this range. An example of this interpolation can be seen in Fig.~\ref{fig:interpolation}, it can be seen that the characteristic shape of the coupling coefficient does not change, and hence we consider this a valid methodology.

\clearpage

\bibliographystyle{apsrev4-1}